\documentclass[graphicx, nofootinbib, notitlepage, twocolumn, amsmath, amssymb, longbibliography]{revtex4-2}
\usepackage{dcolumn}  
\usepackage{bm}       
\usepackage{braket}   
\usepackage{dsfont}
\usepackage{newunicodechar}
\newunicodechar{≤}{\ensuremath{\leq}}
\usepackage{enumitem}
\usepackage{tikz}
\usepackage{amsmath}
\usepackage[small,labelfont=bf,up,textfont=up]{caption} 
\usepackage{caption}
\usepackage{subcaption}
\graphicspath{ {} }
\usepackage[pdftex,bookmarks,breaklinks,bookmarksopen,bookmarksnumbered,linktocpage,pdftex]{hyperref}

\def\nonum{\nonumber \\}
\def\mbf{\mathbf}
\def\t{\text}
\def\mrm{\mathrm}
\def\beit{\begin{itemize}}
\def\eit{\end{itemize}}
\def\mbf{\mathbf}

\def\l{\left}
\def\r{\right}

\def\tb{\textcolor{black}}

\newcolumntype{L}[1]{>{\raggedright\arraybackslash}p{#1}}

\newcolumntype{C}[1]{>{\centering\arraybackslash}p{#1}}

\newcolumntype{R}[1]{>{\raggedleft\arraybackslash}p{#1}}

\begin{document}

\title{Vibrational polariton transport in disordered media}
\author{Enes Suyabatmaz}
\author{Raphael F. Ribeiro}
\email[]{raphael.ribeiro@emory.edu}
\affiliation{Department of Chemistry and Cherry Emerson Center for Scientific Computation, Emory University, Atlanta, GA, 30322}

\date{\today}
\begin{abstract}
Chemical reactions and energy transport phenomena have been experimentally reported to be significantly affected by strong light-matter interactions and vibrational polariton formation. These quasiparticles exhibit nontrivial transport phenomena due to the long-range correlations induced by the photonic system and elastic and inelastic scattering processes driven by matter disorder. In this Article, we employ the Ioffe-Regel criterion to obtain  vibrational polariton mobility edges and to identify distinct regimes of delocalization and transport under variable experimental conditions of light-matter detuning, disorder, and interaction strength. Correlations between the obtained trends and recent observations of polariton effects on reactivity are discussed, and essential differences between transport phenomena in organic electronic exciton and vibrational polaritons are highlighted. Our transport diagrams show the rich diversity of transport phenomena under vibrational strong coupling and indicate that macroscopic delocalization is favored at negative detuning and large light-matter interaction strength. We also find the surprising feature that, despite the presence of dephasing-induced inelastic scattering processes, macroscopic lower polariton delocalization and wave transport are expected to persist experimentally, even in modes with small photonic weight.
\end{abstract}
\maketitle
\section{Introduction}
Electromagnetic field confinement enables strong interactions between molecular excitations and optical microcavity resonances, leading to the formation of hybrid modes denoted molecular polaritons \cite{ebbesen2016, herrera2020, li2022molecular, ribeiro2022introduction, dunkelberger2022vibration}. Recent experiments  have shown that certain polaritonic materials have desirable properties for applications in chemical and materials science, such as controllable charge conductivity \cite{orgiu2015, nagarajan2020},  energy transfer phenomena \cite{coles2014polariton, zhong2017, georgiou2018control, xiang2020intermolecular, son2022energy}, and chemical reactivity \cite{thomas2016, thomas2019, lather2019, thomas2020, hirai2020modulation,  ahn2022modification}. However, the optimal conditions and mechanisms underlying various observations of polariton effects on chemical systems remain unknown \cite{nagarajan2021chemistry, simpkins2021mode, ribeiro2022introduction}. For instance,  experiments operating under strong light-matter interactions between bright normal modes of a 3D molecular ensemble and resonant planar infrared (IR) microcavities have suggested vibrational polariton formation induces cavity-assisted ultrafast intermolecular energy transfer \cite{xiang2020intermolecular} and chemical reactivity modulation  \cite{thomas2016, thomas2019, lather2019, thomas2020, hirai2020modulation,  ahn2022modification}. Theoretical studies \cite{li2021,du2022catalysis, wang2022chemical, lindoy2022quantum} have suggested the reported vibrational strong coupling (VSC) effects on reactivity are caused by anomalous intra or intermolecular vibrational energy redistribution induced by the light-matter coupling, but the key factors determining the intensity and direction of the VSC-induced change on reactivity and energy transfer remain largely unknown \cite{nagarajan2021chemistry, simpkins2021mode, ribeiro2022introduction}. 

\begin{figure}[h]
    \includegraphics[width=\columnwidth]{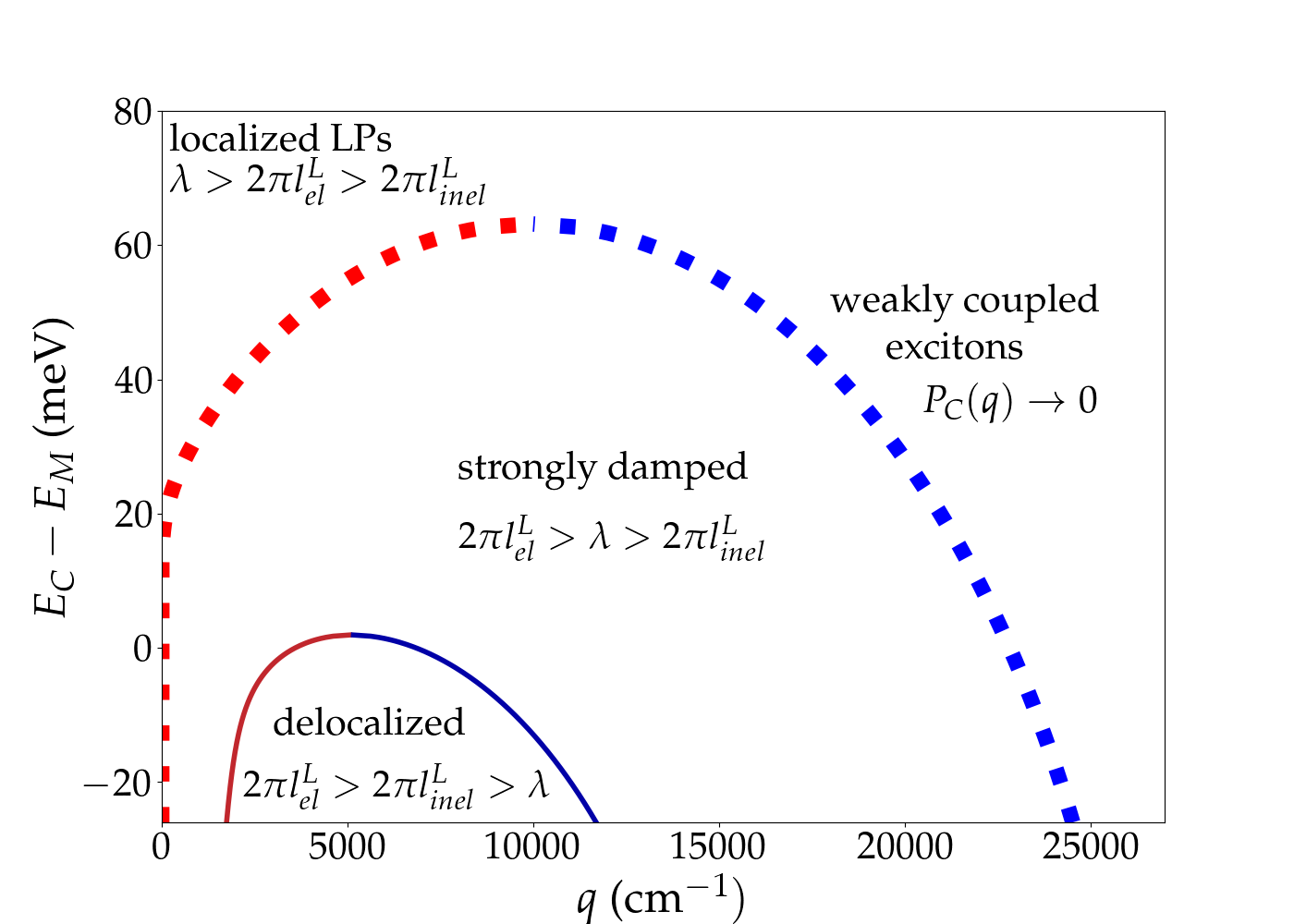}
    \caption{Transport phase diagram for vibrational (lower) polaritons emergent from VSC between an ensemble of $-$C$-$Si of the PFA molecule and an IR microcavity reported in Ref. \cite{thomas2020} with $\hbar\Omega_R = 15.9 ~\t{meV}$, $\sigma_M = 0.967 ~\t{meV}$ and $\gamma_M = 3.97~\t{meV}$ (see Subsec. \ref{ssec:lm_sys_params}). The continuous red and blue curves contain the low and high-energy mobility edges $q_{\t{min,inel}}^{\t{L}}$ and $q_{\t{max,inel}}^{\t{L}}$ for a given detuning. The dotted red and blue curves correspond to the lower and upper boundaries of the delocalization-localization transition in wave number space according to the Ioffe-Regel criterion in the absence of inelastic scattering. Table \ref{tab:classification} provides a description of the transport phenomena expected in each region.}
    \label{fig:thomas_phasediagram3}
\end{figure}

\par Despite the preponderance of questions on the impact of collective (ensemble) vibrational strong coupling on energy transfer in IR microcavities, theoretical and experimental investigations of space-time resolved polariton propagation have been primarily restricted to electronic (organic) exciton-polaritons \cite{pandya2022tuning, xu2022ultrafast, balasubrahmaniyam2023enhanced}. 
\par Polariton propagation and energy transport phenomena in disordered organic electronic exciton-polariton systems was first examined theoretically by Agranovich et al. \cite{agranovich2003cavity} (see also \cite{litinskaya2006}). This pioneering work showed that ballistic, diffusive, and subdiffusive (Anderson localization) propagation are likely to coexist in polaritonic materials due to the large energetic disorder in molecular aggregate samples. Corroboration from numerical simulations was later provided in 1D by Michetti and LaRocca \cite{michetti2005} and Agranovich and Gartstein \cite{agranovich2007}, and experimentally examined in Ref.  \cite{guebrou2012coherent}. Recent experiments have shown additional features of organic exciton-polariton propagation, including quality factor-dependent speed \cite{pandya2022tuning}, tunable energy transport velocity and effective diffusion constant \cite{balasubrahmaniyam2023enhanced}, and superfluid transport under intense  pumping (above the condensation threshold) \cite{lerario2017}.

Collective IR strong coupling shows significant differences relative to organic exciton strong coupling. These differences include the much longer IR microcavity lengths relative \cite{thomas2020, simpkins2015} to the organic, the one to two orders of magnitude weaker collective light-matter interaction strengths achieved with dipole-active high-frequency molecular vibrational transitions relative to the bright electronic excitations of organic materials, and the much narrower vibrational line shapes. The stark contrast in light-matter interaction properties of organic and vibrational excitations suggests the wave functions of vibrational polaritons and their corresponding energy transport processes have qualitatively distinct features relative to those of organic exciton-polaritons. In this work, we present a detailed theoretical analysis of the properties of transport phenomena in disordered systems under VSC.

 We note that several recent studies have investigated polariton and weakly coupled wave functions \cite{scholes2020polaritons,botzung2020a, chavez2021disorder, cohn2022vibrational, gera2022exact, gera2022effects, cederbaum2022cooperative} and energy transfer dynamics \cite{li2021collective, zhou2022interplay, sun2022dynamics, cui2022collective, gomez2023vibrational, perez2023simulating} in disordered 0D (with the optical cavity treated as a zero-dimensional device represented by either one or two isolated boson modes with orthogonal field polarizations) and 1D (with a one-dimensional set of photonic modes) \cite{tichauer2021, allard2022disorder, ribeiro2022multimode,engelhardt2022polarition} environments. However, as is well known from wave function localization theory \cite{anderson1958, abrahams1979scaling,van1999localization,sheng2006introduction}, key qualitative aspects of wave transport are strongly dependent on the number of spatial dimensions and low-dimensional properties are not generalizable  \cite{anderson1958, abrahams1979scaling,sheng2006introduction} (e.g., the tight-binding model with short-range hopping only has a matter-insulator disorder induced quantum phase transition in 3D \cite{abrahams1979scaling,evers2008}). 

In this Article, we adopt the Ioffe-Regel criterion (IRC) to examine vibrational polariton transport phenomena in a 3D disordered medium. We unravel optimal conditions for maximization of the density of macroscopically delocalized polariton modes in a molecular material under VSC with a planar microcavity and present vibrational polariton transport phase diagrams (\tb{see Fig. \ref{fig:thomas_phasediagram3}}) identifying different polariton transport phenomena as a function of cavity-matter detuning, energetic disorder, and molecular dephasing rate for selected experimental systems. Our findings indicate that the density of delocalized polaritons modes is larger for strong coupling between redshifted microcavities and molecular systems with slow dephasing and large collective light-matter interaction strength (Rabi splitting). We also report that the density of macroscopically delocalized polaritons is drastically more impacted by a change in the Rabi splitting than by a change of detuning, and we explore the connection of these trends to recent experimental observations of polariton chemistry dependence on collective light-matter interaction strength. 

This Article is organized as follows. Section \ref{sec:theory} reviews the theoretical framework, including a description of the light-matter Hamiltonian, disorder-induced scattering processes, and the criteria we employed in the classification of polariton wave functions and transport regimes. In Sec. \ref{sec:results_discussion}, we present our main results, including vibrational polariton transport phase diagrams and the density of macroscopically delocalized polariton modes at variable parameters of the light-matter system. A comparison between  transport phenomena in organic electronic exciton and vibrational polaritons is provided in Subsec. \ref{ssec:compare}. We conclude by presenting a summary of our results and a discussion of implications to future polariton chemistry research in Sec. \ref{sec:conclusions}. 

 \section{Theory}\label{sec:theory}
In this section, we explain the theoretical formalism employed to characterize vibrational polariton transport phenomena in macroscopic leaky planar microcavities. Our goal is to construct transport phase diagrams (see Fig. \ref{fig:thomas_phasediagram3} for an example of a detuning$-$wave number diagram) for vibrational polaritons and to identify optimal conditions for macroscopic delocalization under VSC. We employ the IRC \cite{ioffe1960non, van1999localization} to obtain nonperturbative estimates of the low and high-energy boundaries (mobility edges) separating regions in  the (in-plane) wave number space where polaritons show \textit{macroscopic delocalization} from those regions containing localized modes. The latter include weakly coupled molecular states with small photon content as well as low-wave number polariton modes, which are delocalized on a molecular length scale yet have finite localization length \cite{litinskaya2008}. The scattering mean free paths required as input to the IRC are obtained following Litinskaya and Reineker \cite{litinskaya2006}, who applied a method from Lifshitz \cite{lifshitz1987selected} to obtain elastic and inelastic scattering polariton rates from the microscopic light-matter Hamiltonian in a disordered medium.

For the sake of completeness and to introduce our notation, we summarize the employed methodology below. In Subsec. \ref{ssec:hamiltonian}, we introduce the light-matter Hamiltonian and the mean-field solutions employed in Subsec. \ref{ssec:scattering} to obtain polariton scattering mean free paths originating from the various types of disorder-induced scattering treated in this work. The IRC for wave function localization is revisited in Subsec. \ref{ssec:ioffe-regel}, and the classification of polariton transport phases is described in Subsec. \ref{ssec:polariton_phases}. In Subsec. \ref{ssec:density_of_delocalized_modes}, we present the method we applied to estimate the density of delocalized LP modes following the IRC. We conclude the presentation of the theoretical tools used in this work in Subsec. \ref{ssec:lm_sys_params} where we provide some comments on the choice of experimental parameters employed in subsequent quantitative analysis, including a description of how we estimated static disorder parameters for experimental systems that did not include such data.
 
\subsection{Light-Matter Hamiltonian and mean-field solutions}\label{ssec:hamiltonian}
The investigated systems consist of an ideal planar (Fabry-Perot) microcavity under collective strong coupling with a molecular ensemble. The optical cavity is formally infinite and periodic along the in-plane directions $x,y$, and it is finite with length $L_C \in ~O(\mu m)$ along the longitudinal $z$ axis. The molecular material hosted by the photonic device is isotropic, homogeneous, and has significant IR oscillator strength at a transition sampled from a Gaussian probability distribution function. 

 The microcavity resonances are specified by their longitudinal wave vector $k_z(m) =m\pi/L_C, ~ m = \{1,2,...\}$, continuous in-plane wave vector $\mbf{q} = (q_x,q_y)$ corresponding to the projection of the wave momentum along the mirror planes, and polarization $\lambda = \t{TE or TM}$ \cite{kavokin2017microcavities}. The empty cavity Hamiltonian is given by
 \begin{align}
 H_\t{L} = \sum_{m\mbf{q}\lambda} E_C(q,m) a_{m\mbf{q}\lambda}^\dagger a_{m\mbf{q}\lambda},	
 \end{align}
where $q= \sqrt{q_x^2+q_y^2}$, $E_C(q,m) = \hbar\omega_C(q, m)$ and the energy-momentum dispersion of the photonic subsystem is 
\begin{align}\label{eq1}
&E_C(q,m)=E_C\sqrt{1+\l(\frac{qL_{C}}{m\pi}\r)^2}, 
\end{align}
with $E_C \equiv E_C(0,m)=(\hbar c/\sqrt{\epsilon_c}) \times m\pi/L_C$, and $\epsilon_{c}$ is the background dielectric constant. \tb{This microcavity model is approximate, for it assumes perfect mirrors and idealized modes with infinite quality factor. The main effects of cavity decay (induced by coupling to a continuum of free space EM modes \cite{dalton1996field, viviescas2003field, scholes2021emergence}) are incorporated perturbatively in our estimates of polariton inelastic scattering in Subsec. \ref{ssec:scattering}.}

\par The molecular system is treated in the low-energy approximation where only the states (namely, the ground and first excited state) with near-resonant transitions with the photonic resonator are  explicitly treated (the effects of remaining intramolecular degrees of freedom inducing dephasing and polariton decay are effectively accounted by inelastic scattering processes that we discuss later). Intermolecular interactions are ignored since they provide negligible bandwidth to the spectrum of the molecular system in every studied case (as evidenced by the dispersionless molecular absorption spectrum outside an optical microcavity of typical molecular systems in disordered phases). It follows the bare molecule Hamiltonian can be written as
\begin{align}
H_M = \sum_{i=1}^{N_M} \l(\hbar\omega_M + \sigma_i\r) \ket{1_i}\bra{1_i}, \label{eq:hm}
\end{align}
where $\omega_M$ is the mean molecular transition frequency, and $\sigma_i$ is a random variable sampled from a Gaussian probability distribution function with variance $2^{-1/2}\sigma_M$ corresponding to static disorder induced on the molecular normal-modes by slow system-environment (solvent) correlations on the transport timescales relevant to us. Note that our work is focused on vibrational polariton transport in the dilute quasiparticle regime where the polariton density is always small enough that nonlinearities of the light-matter Hamiltonian can be ignored.

In all investigated scenarios, we examine the polaritons originating from the strong interaction of a single microcavity band with fixed $k_z \equiv k_z(m)$ and the near-resonant molecular ensemble. The effects of other photonic modes will be ignored as they are either highly off-resonant with the molecular material, or in cases where the molecular resonance overlaps more than a single photon band, the corresponding resonance in-plane wave vectors are far enough from each other that the polaritons resulting from distinct photonic bands may be treated independently.

The light-matter interaction is treated in the Coulomb gauge within the rotating-wave-approximation
\small
\begin{align}
H_{\t{LM}} =\sum_{m\mbf{q}\lambda} \sum_{n=1}^{N_M}  g_{m\mbf{q}\lambda}(\mbf{R}_n) a_{m\mbf{q}\lambda}\sigma_n^+ + g_{m\mbf{q}\lambda}^*(\mbf{R}_n) a_{m\mbf{q}\lambda}^\dagger \sigma_n^-, \label{eq:hlm}
\end{align}
\normalsize
where $\sigma_n^+ = \ket{1_n}\bra{G}$ and $\sigma_n^- = \ket{G}\bra{1_n}$ and $\ket{G}$ is the vacuum state of the light-matter system where all microcavity modes and (relevant) molecular degrees of freedom are in their  ground-state. Assuming the electromagnetic environment is generated by a perfect planar microcavity, the interaction strength $g_{m\mbf{q}\lambda}(\mbf{R}_n)$ between molecule $n$ (treated as a point-dipole) located at $\mbf{R}_n = (x_n, y_n, z_n)$ and the $(m\mbf{q}\lambda)$ mode of the EM field is given in the Coulomb gauge by
\small
\begin{align}
& g_{m\mbf{q}\text{TE}}(\mbf{R}_n) = -i \frac{E_M} {E_C(q)}\sqrt{\frac{E_C(q)}{2\epsilon}} \frac{e^{i\mbf{q}\cdot \mbf{R}_n}}{\sqrt{S}} \mu_n \cdot \mbf{f}_{m\mbf{q}\t{TE}}(\mbf{R}_n), \label{eq:coupling1} \\
&g_{m\mbf{q}\t{TM}}(\mbf{R}_n) = -i\frac{E_M} {E_C(q)}\sqrt{\frac{E_C(q)}{2\epsilon}} \frac{e^{i\mbf{q}\cdot \mbf{R}_n}}{\sqrt{S}} \mu_n\cdot \mbf{f}_{m\mbf{q}\t{TM}}(\mbf{R}_n),
\end{align}\normalsize
with mode profile vectors $
\mbf{f}_{m\mbf{q}\t{TE}}(\mbf{R})$ and $\mbf{f}_{m\mbf{q}\t{TM}}(\mbf{R})$ given by \cite{jackson2021classical,zoubi2005microscopic} \small
\begin{align}
	& \mbf{f}_{m\mbf{q}\t{TE}}(\mbf{R}) = \sqrt{\frac{2}{L_C}}\t{sin}\l(k_z z\r) \mbf{e}_z \times \mbf{e}_\mbf{q}, \\
	&\mbf{f}_{m\mbf{q}\t{TM}}(\mbf{R}) = \sqrt{\frac{2}{L_C}} \frac{k^2}{k_z^2} \l[\t{sin}\l(k_z z\r)\mbf{e}_\mbf{q}  -\frac{iq}{k_z z}\t{cos}(k_z z) \mbf{e}_z\r],
\end{align}
\normalsize
where $k^2 = k_z^2 + q^2$, \textcolor{black}{$\mbf{e}_z$ is the unit vector along the $z$ axis, $\mbf{e}_{\mbf{q}} = (q_x,q_y)/q$}, the molecular transition-dipole moments $\mu_n$ are uncorrelated random vectors sampled from a uniform distribution on a circle along the $xy$ plane (for simplicity since this assumption is inconsequential), and the molecular positions $\mbf{R}_n$ are uniformly distributed with a mean distance between neighboring molecules equal to $a \equiv \rho^{-1/3}$ where $\rho$ is the molecular density.

The disorder makes it challenging to extract the exact spectrum of the total light-matter Hamiltonian 
\begin{align}
H = H_\t{M} + H_\t{L} + H_{\t{LM}},	\label{eq:htotal}
\end{align}
due to the large number of coupled light and matter degrees of freedom. However, approximate solutions can be obtained in the weak disorder limit where elastic and inelastic scattering processes are irrelevant, leading to the mean-field lower polariton (LP) and upper polariton (UP) energies $E_{\t{L}}^{(0)}(q)$ and $E_{\t{U}}^{(0)}(q)$ \cite{litinskaya2009gap, agranovich2011hybrid}  
\small
\begin{align}
    E_{\t{U,L}}^{(0)}(q)=\frac{E_{M}+E_{C}(q,m)}{2}\pm \frac{1}{2}\sqrt{\l(\hbar\Omega_R\r)^2 + \l[E_{M}-E_{C}(q,m)\r]^2}, \label{eq:pol_energies}
\end{align}
\normalsize
where $\Omega_R$ is the Rabi splitting frequency (collective light-matter interaction strength). The photonic and molecular weights of the mean-field LP and UP solutions can be straightforwardly obtained from the Schrodinger equation using the same assumptions employed to obtain Eq. \ref{eq:pol_energies} \cite{litinskaya2006, agranovich2009excitations}. \tb{While the polariton energies obtained from Eq. \ref{eq:pol_energies} resemble those given by the Tavis-Cummings model \cite{tavis1968,tavisApproximateSolutionsNMoleculeRadiationField1969}, this occurs here solely because,  in the mean-field limit, Eq. \ref{eq:htotal} can be decomposed into a direct sum of $2\times 2$ Hamiltonians describing the interaction of the cavity mode $(m, \mbf{q},\lambda)$ with a collective matter polarization mode with same symmetry \cite{vinogradov1992vibrational, litinskaya2006}. Nevertheless, our treatment of vibrational polariton propagation below relies fundamentally on the interaction of a common set of molecules with multiple EM modes}. The mean-field polariton eigenvalues and eigenfunctions provide the starting point for the treatment of polariton transport in disordered media described in the next subsections.

\subsection{Scattering mechanisms and mean free paths}\label{ssec:scattering}
 Molecular energetic, dipolar, and translational disorder and system-environment interactions may cause polariton scattering, randomization of their in-plane wave vector, and incoherent transitions between polaritons and reservoir modes \cite{agranovich2003cavity, litinskaya2006, litinskaya2008}. Elastic scattering induced by nonvanishing static energetic disorder and structural medium fluctuations preserve the wave vector magnitude $q$ but randomizes its direction, whereas inelastic scattering (photon leakage or via molecular system-bath interactions) also dissipates energy from the polariton system and therefore does not conserve $q$. 
 
 The average distance between elastic [inelastic] scattering events for a wave packet with mean momentum $q$ in the weak scattering limit is given by the mean free path $l_{\t{el}}(q)$ [$l_{\t{inel}}(q)$]. As discussed below, the mean free paths associated with distinct scattering mechanisms may be employed to characterize \cite{van1999localization, sheng2006introduction, litinskaya2008, skipetrov2018, poduval2023anderson} the transport properties of polariton quasiparticles.
 
The mean polariton free path $l_s^{\t{P}}(q)$ ($\t{P} = \t{LP}$ or $\t{UP}$), associated with the scattering mechanisms $s = $ inelastic, resonance (elastic), and induced by fluctuations of molecular  orientation and positions (elastic), can be obtained from the corresponding scattering rates $\Gamma_s^{\t{P}}(q)/\hbar$ and the (mean-field) polariton group velocity $v_{\t{P}}(q)$ via 
\begin{align} 
&l_s^{\t{P}}(q)= \frac{\hbar v_\t{P}(q)}{\Gamma_s^{\t{P}}(q)}, \label{eq:mfp}\\
 &   v_\t{P}(q)= \frac{1}{\hbar}\frac{\partial E_\t{P}^{(0)}(q)}{ \partial q} = \l(\frac{L_C}{m\pi}\r)^2 \frac{q E_{C}^{2}}{\hbar E_{C}(q,m)}P_C^{\t{P}}(q),
\end{align}	
where $P_C^{\t{P}}(q)$ is the photon weight in the corresponding (mean-field) polariton mode. The wave-vector broadening induced by the scattering process $s$ on the polariton mode $\t{P}(q)$ is given by  $\delta q_{s}^\t{P}(q)= \l[l_s^{\t{P}}(q)\r]^{-1}$ \cite{sheng2006introduction}. 

Elastic scattering can be induced by static (diagonal) energetic disorder (Eq. \ref{eq:hm}) and structural fluctuations of the environment (off-diagonal disorder, Eqs. \ref{eq:hlm}). As we show below, elastic scattering produced by the inhomogeneous distribution of molecular transition energies is dominant. This process, typically denoted by resonance scattering \cite{tiggelen1991, litinskaya2006} occurs when a polariton with wave vector $\mbf{q}$ is converted into another with wave vector $\mbf{q}'$ with $q' = q$, via an intermediate nearly resonant weakly coupled molecular excited-state. Resonance scattering is often considered to be weak relative to absorptive processes (e.g., ``decay into dark modes"). However, as shown by Agranovich et al. \cite{agranovich1987}, and Litinskaya et al. \cite{litinskaia2002elastic} in their earlier studies of inorganic microcavities, resonance scattering can dominate over inelastic processes under experimentally achievable scenarios where the energy scale of polariton inelastic scattering is smaller than the energetic disorder. 

When the density of states of the bare molecular system is significant at $E_\t{P}^{(0)}(q)$, resonance scattering is expected to be strong, and a perturbative treatment is invalid. However, a nonperturbative estimate of the scattering rate $\Gamma_{\t{res}}^{\t{P}}(q)$ may be obtained as the imaginary part of the complex polariton energies $E_{\t{LP}}(q)$ or $E_{\t{UP}}(q)$ arising as solutions to the complex eigenvalue equation (originating from Eq. \ref{eq:htotal} after assuming scattering induced by structural fluctuations can be ignored \cite{litinskaya2006}) \begin{align}
   & E(q)-E_{C}(q,m)=\frac{\l(\hbar\Omega_R\r)^2}{4} \int_{-\infty}^{\infty}dE_j \frac{\rho(E_j)}{E(q)-E_j}
    \nonum
   &=\frac{\l(\hbar\Omega_R\r)^2\sqrt{\pi}}{4i\sigma}e^{-[E(q)-E_M]^2/\sigma^2}\text{erfc}[-i[E(q)-E_M]/\sigma] \label{eq:res_scatter},
\end{align}
where $\rho(E_j)=(1/\sigma_M\sqrt\pi)\t{exp}\l[-(E_j-E_M)^2/\sigma_M^2\r]$, and $\text{erfc}$ is the complementary complex error function. This equation 
can be solved numerically to find  $\Gamma_{\t{res}}^{\t{P}}(q)$ as $\t{Im}\l[E_{\t{P}}(q)\r] = \Gamma_{\t{res}}^{\t{P}}(q)$ for application in Eq. \ref{eq:mfp} providing the resonant scattering mean free path for polariton mode $P$ with wave number $q$.

Fourth-order perturbation theory in structural deviations from the mean \cite{litinskaya2006} provides an estimate of the mean free path for scattering induced by medium fluctuations  from the corresponding elastic scattering rate
\begin{equation}
   \Gamma_{\t{fluc}}^{\t{P}}(q)\approx\frac{3m^2\pi^2}{2}\l(\frac{a}{L_C}\r)^3\frac{\l(\hbar\Omega_R\r)^2}{4E_C}P_M^{\t{P}}(q),
    \label{eq:fl_scatter}
\end{equation}
where $P_M^\t{P}(q)$ denotes the total molecular probability in the polariton mode $\t{P}(q)$. Note that in VSC of small and medium-sized molecules with Fabry-Perot cavities, the intermolecular distance $a$ is almost always on the order of a few to tens of nanometers, whereas $L_C = O\l(10^3-10^4 ~ \t{nm}\r)$, and $\hbar\Omega_R \ll E_C$. Therefore, $\Gamma_{\t{fluc}}$ will almost always be negligible relative to resonance elastic scattering and inelastic decay, especially at $q \gg 0$.

In the absence of strong interactions between the normal-modes strongly coupled to the microcavity, inelastic scattering processes induce polariton decay via at least two mechanisms: (i) local interactions between the strongly coupled molecular states and intra or intermolecular \tb{(e.g., low-frequency solute-solvent)} bath modes which induce polariton dephasing (via decay into weakly coupled or dark modes)\cite{litinskaya2008, pino2015}, and (ii) photon escape via leaky cavity mirrors \cite{kavokin2017microcavities}. Both processes extinguish polaritons while transferring energy irreversibly into a thermal reservoir (molecular and photonic, respectively) with a significantly larger density of states. The rates of both types of events are given by  $\gamma_{M}/\hbar$ and $\gamma_{C}/\hbar$, respectively. The cavity leakage rate $\gamma_C$ can be obtained from the bare cavity mode linewidth or quality factor \cite{kavokin2017microcavities, steck2017a} $Q = \hbar\omega_C/\gamma_C$, whereas $\gamma_M$ \tb{can be obtained from the molecular homogeneous absorption linewidth}. In this work, we obtain an approximation for $\gamma_M$ from the polariton linewidth at resonance \cite{houdre1996} (see Subsec. \ref{ssec:lm_sys_params}). \tb{Recently, Li et al. reported an additional contribution to polariton inelastic scattering via direct interactions between nearest neighbor molecular modes contributing to polaritons \cite{li2022polariton}. In Appendix B, we provide an approximate upper bound to this intermolecular dephasing mechanism for systems where the solute of a mixture undergoes VSC (like those we examined in this work, see Table \ref{tab:parameters}), numerically compare it to Eq. \ref{eq:gamma_inel}, and show that the processes (i) and (ii) listed above provide the dominant inelastic scattering pathways (we also note that, in scenarios where strong coupling occurs with neat liquids or molecular crystals, the dephasing pathway introduced in Ref. \cite{li2022polariton} likely plays a more important role).}

\par \tb{Assuming $\gamma_M$ is dominated by dephasing induced by interactions of each molecular normal-mode with a low-frequency bath}, the total inelastic scattering rate for a polariton with in-plane momentum $q$ can be estimated by \tb{(see Appendix A for a derivation)}
\begin{equation}
    \Gamma_{\t{inel}}^{\t{P}}(q)= P_M^{\t{P}}(q)\gamma_{M}+P_C^{\t{P}}(q)\gamma_{C}. \label{eq:gamma_inel}
\end{equation}
Note that, in principle, $\Gamma_{\t{inel}}^{\t{P}}(q)$ includes both processes that conserve the polariton-number [e.g., $\t{P}(q) \rightarrow \t{P}'(q')$ with $q' \neq q$ and $\t{P}$ and $\t{P}'$ refer to either LP or UP], and processes where a polariton decays into a weakly coupled molecular state or decays via emission of a photon bringing the system into the ground-state. The processes annihilating polaritons are expected to be dominant in collective strong coupling measurements due to the large density of weakly coupled molecular modes and ultrafast photon leakage via highly leaky mirrors. Regardless of their outcomes, inelastic processes are induced by interaction with an incoherent reservoir. Therefore, the inelastic scattering mean free path may also be viewed as the polariton phase-breaking length since it provides the average distance traveled by a wavepacket with mean wave number $q$ before the phase relations between its components are randomized. 

\subsection{Ioffe-Regel criterion for localization}\label{ssec:ioffe-regel}
Using the scattering mean free paths associated with inelastic, resonant, and fluctuation-induced elastic scattering obtained from the expressions above, the regions in the $q$ space where vibrational polariton wave functions are macroscopically delocalized (can transport energy over arbitrarily large distances) can be determined via the Ioffe-Regel criterion \cite{ioffe1960non, van1999localization}. 

According to the IRC, polariton localization occurs when the (in-plane) wavelength approaches its (shortest) mean free path. Specifically, the wave number space boundaries between macroscopically delocalized and localized polariton modes are obtained from the condition
 \begin{align} q l^\t{P}(q)  = \frac{2\pi l^{\t{P}}(q)}{\lambda(q)}\approx 1, \label{eq:irc} \end{align} 
 where $l^{\t{P}}(q)$ is the mean free path corresponding to the strongest scattering process perturbing $\t{P} = $ LP or UP with wave vector $\mbf{q}$ and $\lambda(q)$ is the in-plane wavelength. 
 
 For a given set of parameters characterizing the light-matter system (detuning $u = E_C - E_M$, Rabi splitting $\Omega_R$, static disorder $\sigma_M$, cavity leakage rate, $\gamma_C$, and molecular homogeneous line width $\gamma_M$), the wave number boundaries $q_{\t{min}}$ and $q_{\t{max}}$ separating localized polariton modes ($q > q_{\t{max}}$ and $q < q_{\t{min}}$) from delocalized ($ q_{\t{min}} < q < q_{\t{max}}$) determine the mobility edges for polariton transport. The IRC for wave localization can also be equivalently written as $q/\delta q < 1$ in terms of the mean in-plane wave number fluctuation $\delta q = 1/l^{\t{P}}(q)$.

Notably, while the IRC is understood to provide only an approximate estimate for the mobility edges of a disordered system, it has numerical support from multiple works \cite{litinskaia2002elastic, skipetrov2014, skipetrov2018} and may be justified formally \cite{tiggelen1991, economou1984localized} using the scaling theory of localization \cite{abrahams1979scaling, sheng2006introduction} and the self-consistent theory of localization \cite{vollhardt1980diagrammatic, sheng2006introduction}. 

To find the LP and UP mobility edges for a given light-matter system, we first (numerically) solve the independent IRC equations associated with each scattering mechanism $s$  \begin{align}
	q l^\t{P}_s(q)  = 1, ~~ s \in \{\t{fluc, res, inel}\}. \label{eq:irc_s}
\end{align}
The set of solutions of this equation may have 0, 1, or 2 elements depending on $s$ and the parameters of the light-matter system. For instance, the resonance scattering LP mean free path is, in general, a nonmonotonic function of $q$ at zero or negative detuning. Therefore, the generic set of solutions of Eq. \ref{eq:irc_s} for LP resonance scattering is given by $Q_{\t{res}}^{\t{LP}} \equiv \{q_{\t{min,res}}^\t{L},q_{\t{max,res}}^\t{L} \}$, where min and max refer to the corresponding minimum and maximum solutions. This implies that in the absence of inelastic and fluctuation scattering, only LP modes with $q_{\t{min,res}}^{\t{L}} < q < q_{\t{max,res}}^{\t{L}}$ would show macroscopic delocalization.  

In comparison to resonance scattering, inelastic processes have qualitatively distinct effects on polariton wave packet evolution that we have anticipated above and will further discuss below. Nevertheless, inelastic mean free paths also show, in general, nonmonotonic behavior with $q$ (Fig. \ref{fig:mfp_thomas}). We denote the associated typical solution set of Eq. \ref{eq:irc_s} by $ Q_{\t{inel}}^{\t{L}} \equiv \{q_{\t{min,inel}}^\t{L},q_{\t{max,inel}}^\t{L} \}$. In principle, these solutions carry the same interpretation as the resonant case discussed above: only LP modes with $q_{\t{min,inel}}^\t{L}<q <  q_{\t{max,inel}}^\t{L}$ are delocalized over macroscopic distances according to the IRC.

Conversely, for the systems examined in this article, the LP fluctuation mean free path is always macroscopic (Fig. \ref{fig:mfp_thomas}), and the generic solution set of Eq. \ref{eq:irc_s} for $s = \t{fluc}$ has, a single element, $Q_{\t{fluc}}^{\t{L}} \equiv \{ q_{\t{fluc}}^{\t{L}} \}$. This implies that in the absence of energetic disorder and inelastic scattering, all LP modes with $q > q_{\t{fluc}}^{\t{L}}$ are delocalized. 

The sets $Q_{\t{fluc}}^{\t{L}}, Q_{\t{res}}^{\t{L}}$ and $Q_{\t{inel}}^{\t{L}}$ contain all information required to construct the LP transport phase diagrams. Strictly speaking, long-lived propagating polariton modes must satisfy $ql_s^\t{P}(q) > 1$ for each type of scattering $s$. Therefore, the minimum $q$ (low-energy mobility edge) is determined by the scattering process $s$ with a maximum value of $q_{\t{min},s}$, since $q l_s^{\t{L}}(q) < 1$ for all $q < q_{\t{min},s}$ implying localized polariton modes according to the IRC. On the other hand, the high-energy mobility edge is set by the scattering process with a minimum value of $\{q_{\t{max},s}\}$, since strong localization of the polariton wave function happens for $q> q_{\t{max}}$ regardless of the effectiveness of other scattering mechanisms. To summarize, the minimum and maximum in-plane wave numbers (low and high-energy mobility edges) for macroscopically delocalized LP modes are given in general by
\begin{align}
& q_{\t{min}}^{\t{L}} = \t{max}\l\{q_{\t{min,res}}^{\t{L}}, q_{\t{min,fluc}}^{\t{L}}, q_{\t{min,inel}}^\t{L}\r\}, \label{eq:qminl} \\ 
& q_{\t{max}}^\t{L} = \t{min}\l\{q_{\t{max,res}}^\t{L}, q_{\t{max,inel}}^\t{L}\r\} \label{eq:qmaxl}.
\end{align}

A similar analysis can be performed for UP, but in that case, we only find a low-energy mobility edge as Eq. \ref{eq:irc_s} admits, in general, a single solution $q_{\t{min}}^{\t{U}}$. All modes with $q < q_{\t{min}}^{\t{U}}$ are localized, while UP states with $q > q_{\t{min}}^\t{U}$ display macroscopic delocalization.  This occurs even in redshifted microcavities since only a small fraction of UP modes have significant molecular content. Note that for sufficiently large $q$, UP modes may be regarded as photons weakly perturbed by off-resonant molecular scattering, see, e.g., Fig. \ref{fig:dispersion_curves_thomas}). Therefore, from now on, we will focus our analysis of transport via the LP modes emergent from VSC.
\renewcommand{\arraystretch}{1.25}
\begin{widetext}
\begin{table*}[t]
\caption{Summary and description of polariton classification criteria adopted in this work. Below and elsewhere in the text, $l_{\t{el}}^{\t{L}}$ refers to the shortest elastic mean free path among resonance and fluctuation-induced elastic scattering. Likewise, $q_{\t{min,el}} ^{\t{L}}$ and $q_{\t{max, el}}^{\t{L}}$ refer to the solutions of Eqs. \ref{eq:qminl} and \ref{eq:qmaxl} in the absence of inelastic scattering (i.e., when $q_{\t{min,inel}} = 0$ and $q_{\t{max,inel}} \rightarrow \infty$).}
\begin{tabular}{L{3cm}C{4cm}L{9cm}}
\hline 
\textbf{Mode character} & \textbf{Criterion} & \textbf{Description} \\
\hline
delocalized &  $\lambda < 2\pi l_{s}^{\t{L}} $ &  Delocalized modes are the optimal energy transport carriers. \tb{In other words, wave packets consisting of a narrow superposition of polariton modes with wave vector centered around $\mbf{q}$ with $q l_{s}^{\t{L}}(q) > 1$ travel ballistically with a group velocity $|\nabla_\mbf{q} \omega_P(\mbf{q})|$ prior to elastic and inelastic scattering times \cite{allen1998evolution, sheng2006introduction}}. However, delocalized modes can behave in significantly different fashions depending on their main scattering mechanism. When elastic scattering prevails over inelastic collisions, interference effects lead to weak localization, i.e., a reduced diffusion constant and enhanced coherent backscattering \cite{sheng2006introduction}. The classical diffusive regime (with decay) for energy transport arises only when incoherent inelastic scattering that conserves the polariton number occurs on a faster timescale than elastic ($l_{\t{inel}} < l_{\t{el}}$).\\ 
strongly damped propagating &  $2\pi l_{\t{el}}^{\t{L}} > \lambda > 2\pi l_{\t{inel}}^{\t{L}}$ & Assuming that the dominant effect of inelastic scattering is to induce polariton light emission or molecular absorption via dephasing, strongly damped propagating modes are expected to transport energy with weak elastic scattering within their lifetimes. 
  \\
localized LPs & $\lambda > 2\pi l_{\t{el}}^{\t{L}}$ and $q < q_{\t{min,el}}^{\t{L}}$ &  Localized LPs satisfy the IRC for localization and have small enough $q$ that their photon content is significant (typically the case for redshifted or near-resonant cavities). Conventional kinetic theory fails for these modes. If a localized wave packet is launched with mean in-plane wave number $q$ satisfying the criterion given here, it is expected to show transient diffusive behavior and localization. If inelastic scattering dominates, then it is likely that the wave packet will decay before showing signatures of strong localization. \\
(localized) weakly coupled excitons  & $\lambda > 2\pi l_{\t{el}}^{\t{L}}$ and $q > q_{\t{max}}^{\t{L}}$ & The majority of these states has less than 1$\%$ photonic content. Note that there is no sharp boundary between localized LP and weakly coupled molecular modes (Fig. \ref{fig:thomas_phasediagram3}). However, we make a distinction between these two types of localized states because LP modes with $q \rightarrow 0$ are expected to show a delocalization length that is large on the molecular length scale (as verified by numerical simulations \cite{agranovich2007, michetti2005}), whereas weakly coupled molecular modes are expected to localize over much smaller distances \cite{ribeiro2022multimode, allard2022disorder, engelhardt2022polarition}.\\
\hline
\label{tab:classification}
\end{tabular}
\end{table*}%
\end{widetext}
\subsection{Polariton transport phases}\label{ssec:polariton_phases}
The IRC allows us to distinguish macroscopically delocalized polaritons from strongly localized modes. This classification may be further refined by recognizing two essential aspects of polariton kinetics.

First, while modes with $q < q_{\t{min}}^{\t{L}}$ and $q > q_{\t{max}}^{\t{L}}$ are both strongly localized according to the IRC, their localization length can differ by many orders of magnitude \cite{michetti2005,agranovich2007, litinskaya2008}. In particular, LP modes with $q < q_{\t{min}}^{\t{L}}$ have significant photon content and are expected to be localized over length scales that are orders of magnitude larger than LP modes with insignificant photonic weight and $q \gg q_{\t{max}}^{\t{L}}$. Therefore, in the phase diagrams we construct here, we make a distinction between localized LP modes and weakly coupled excitons. Both types of modes are strongly localized in the macroscopic sense and belong to the LP branch of the polariton dispersion, but the vast majority of the weakly coupled excitons have almost insignificant photonic content, whereas the inverse is true for localized LP modes. Note this distinction is also suggested by Agranovich \cite{agranovich2009excitations} and Litinskaya \cite{litinskaya2008} and has also been highlighted in recent experimental work \cite{myers2018polariton}.
 
 Second, we denote by ``overdamped propagating modes", polariton states which are only considered to be localized according to IRC due to strong inelastic scattering. The corresponding modes would be delocalized in the absence of molecular dephasing ($\gamma_M = 0$) and photon leakage ($\gamma_C = 0$). Given that polariton decay by cavity leakage or via dephasing into the weakly coupled reservoir are the dominant inelastic scattering processes, we expect that overdamped propagating modes can transport energy efficiently during their lifetime. In other words, polariton wave function strong localization is unlikely to be a generic feature of overdamped propagating modes, although more work is needed to rigorously characterize the transport properties of these states. A summary of our classification scheme is given in Table \ref{tab:classification}.
\subsection{Density of delocalized LP modes}\label{ssec:density_of_delocalized_modes} For the various light-matter systems investigated in this paper, we also analyze under various conditions the total density of delocalized LP modes $\rho_{L}^{\t{deloc}}$. This quantity provides a simple global measure of delocalization in a polaritonic system. Given $q_{\t{min}}^{\t{L}}$ and $q_{\t{max}}^{\t{L}}$, we find $\rho^{\t{deloc}}_{L}$ from
\begin{equation}
  \rho^{\t{deloc}}_{L}  \equiv \frac{1}{2\pi} \int_{q_\t{min}^L}^{q_{\t{max}}^L}\mrm{d}q~q  = \frac{\l(q_{\t{max}}^{\t{L}}\r)^2 - \l(q_{\t{min}}^{\t{L}}\r)^2}{4\pi}.
\end{equation}
Note that we only report the density of delocalized LP modes since the corresponding density of delocalized UP modes would be dependent on the choice of high-energy cutoff momentum. In any case, the fraction of delocalized UP modes is expected to be close to one since only a  small fraction of UPs (with generally near zero in-plane wave number) have significant molecular content and may become localized (see, e.g., Fig. \ref{fig:dispersion_curves_thomas}).
\subsection{Light-matter systems and parameters}\label{ssec:lm_sys_params}
The set of parameters employed to obtain the vibrational polariton phase diagrams presented in this Article are listed in Table \ref{tab:parameters}. All input parameters were either obtained directly or derived from experimental data in the cited references. The molecular energetic disorder parameters $\sigma_M$ for the systems in Refs. \cite{thomas2020, lerario2017} were estimated in the following way: the relevant bare molecule excitations were assumed to have a Voigt line shape \cite{wertheim1974determination} with Lorentzian FWHM $\gamma_M$ and Gaussian inhomogeneous broadening $\sigma_M$. The parameter $\gamma_M$  was obtained from the (approximate) LP line width $\gamma_\t{LP} = (\gamma_M + \gamma_C)/2$  at resonance \cite{houdre1996}. Using the experimental value for the Voigt line width $f_V$ and the approximate form \cite{whiting1968empirical} 
\begin{align} f_V \approx \frac{\gamma_M}{2} + \sqrt{\frac{\gamma_M^2}{4}+ 4\sigma_M^2 \t{ln}(2)}, \end{align} 
we  solved for $\sigma_M$ in each case to obtain the values reported in Table \ref{tab:parameters}.

\section{Results and Discussion}\label{sec:results_discussion}
We provide transport phase diagrams and the density of delocalized polariton modes for several systems under vibrational strong coupling in this section. Our focus is on IR polariton energy transport, but we also provide a comparison to a recently studied electronic (organic) exciton-polariton system below. Table \ref{tab:parameters} includes the input parameters, LP mobility edges, and density of delocalized LP modes for selected systems.

In Subsec. \ref{ssec:detuning}, we begin our analysis of vibrational polariton transport by investigating $q-u~(\t{wave number} $-$\t{detuning}$) transport phase diagrams obtained with parameters representing the vibrational strong coupling between the $m=2$ band of an IR microcavity with a normal-mode involving a significant contribution from the $\t{C}-\t{Si}$ bond of 1-phenyl-2-trimethylsilylacetylene (PTA) studied recently by  Thomas et al. \cite{thomas2020}. This study established that an earlier reported VSC effect on chemical reactivity \cite{thomas2016} can be controlled by varying the light-matter interaction strength via manipulation of the molecular density. 

Subsequently, we analyze how the previously obtained detuning-dependent transport phase diagram is modified in the limit where molecular dynamical disorder (here represented by the bare molecule dephasing rate $\gamma_M$; Inh. lim. case in Table \ref{tab:parameters}) vanishes and the bare molecule line width is entirely due to inhomogeneous broadening, where $\sigma_M$ takes its maximal value (assuming a Voigt line shape as explained in Sec. \ref{ssec:lm_sys_params}). This discussion is followed with an exploration of the transport phase diagram obtained in the homogeneous broadening limit where $\sigma_M = 0$ and $\gamma_M$ is maximized (Hom. lim. in Table \ref{tab:parameters}). 

Subsection \ref{ssec:disorder} examines the effects of variable disorder on
polariton transport phenomena. For a fixed  (zero) detuning, we establish the universal forms of the $q-\sigma_M$ and $q-\gamma_M$ transport phase diagrams providing the critical disorder and dephasing rates beyond which no macroscopic delocalized LP mode exists according to the IRC criterion.

In Subsection \ref{ssec:rabi}, we present an analysis of the effect of the collective light-matter interaction strength on the mobility edges and density of delocalized polariton modes at zero detuning and $\sigma_M$ and $\gamma_M$ fixed according to the experimental data of Ref. \cite{thomas2020}.

We conclude this section with a comparison of the main features of organic exciton and vibrational polariton by comparing their (detuning dependent) transport phase diagrams (Subsec. \ref{ssec:compare}).

\renewcommand{\arraystretch}{1.25}
\begin{widetext}
\begin{table*}
\caption{Input parameters, low and high-energy LP mobility edges (minimum and maximum boundaries, $q_{\text{min}}$  and $q_{\text{max}}$ respectively), and density of delocalized LP modes  $\rho_L^{\t{deloc}}\l(\t{cm}^{-2}\r)$ for the infrared and organic polariton systems examined in this work. Energies are given in meV, and wave vector magnitudes are given in cm$^{-1}$. Inh. and Hom. refer respectively to inhomogeneous and homogeneous limits of the bare molecule line shape employed for theoretical analysis of systems with all parameters corresponding to a particular experiment except for the energetic disorder $\sigma_M$ and dephasing rate $\gamma_M/\hbar$ which take the values listed on the Table and explained in Sec. \ref{ssec:detuning}. N/A means not available.}
\begin{tabular}[b]{lccccccccccccc}
\hline 
&$E_{C}$ & $E_M$ & $\hbar\Omega_{R}$ & $\sigma_M/\hbar\Omega_R$ & $a~(\text{nm})$ & $L_C~(\mu$m) &$\hbar\gamma_M$ & $\hbar\gamma_C$ & $q_{\text{min}}^\t{U}$ & $q_{\text{min}}^\t{L}$ & $q_{\text{max}}^{\t{L}}$ &  $P_C\l(q_{\t{max}}^\t{L}\r)$ & $\rho_L^{\t{deloc}} $\\
\hline
$-$C$-$Si (PFA) ~\cite{thomas2020}& 106.6 & 106.6 & 15.87 & 0.06 & 0.69 & 6.0 & 3.97 & 3.47 & 2551 & 3595 & 6843 & 0.103 & $2.7 \times 10^6 $\\
$-$C$-$Si (PFA) (Inh.) & 106.6 & 106.6 & 15.87 & 0.18 & 0.69 & 6.0 & 0.00 & 3.47 & 1904 & 1904 & 11592 & 0.021 & $1.0 \times 10^{7}$ \\
$-$C$-$Si (PFA) (Hom.) & 106.6 & 106.6 & 15.87 & 0.00 & 0.69 & 6.0 & 4.84 & 3.47 & 2654 &  $-$ & $-$ & $-$ & 0.00\\
$-$NCO stretch (HMDI) \cite{simpkins2015} & 269 & 281 & 13.2 & 0.16 & N/A & 1.6 & 3.31 & 6.2 & 4437 & 3237 & 11257 & 0.045 & $9.3 \times 10^6 $\\
TDAF exciton \cite{lerario2017} & 3045 & 3500 & 600 & 0.52 & 1.13 & 0.13 & 18 & 5.0 & 93276 & 13501 & 234471 & 0.111 & $4.36 \times 10^{9}$\\
\hline
\label{tab:parameters}
\end{tabular}
\end{table*}%
\end{widetext}

\subsection{Detuning dependent vibrational polariton transport phase diagrams}\label{ssec:detuning}

\begin{figure}[h]
\includegraphics[width=\columnwidth]{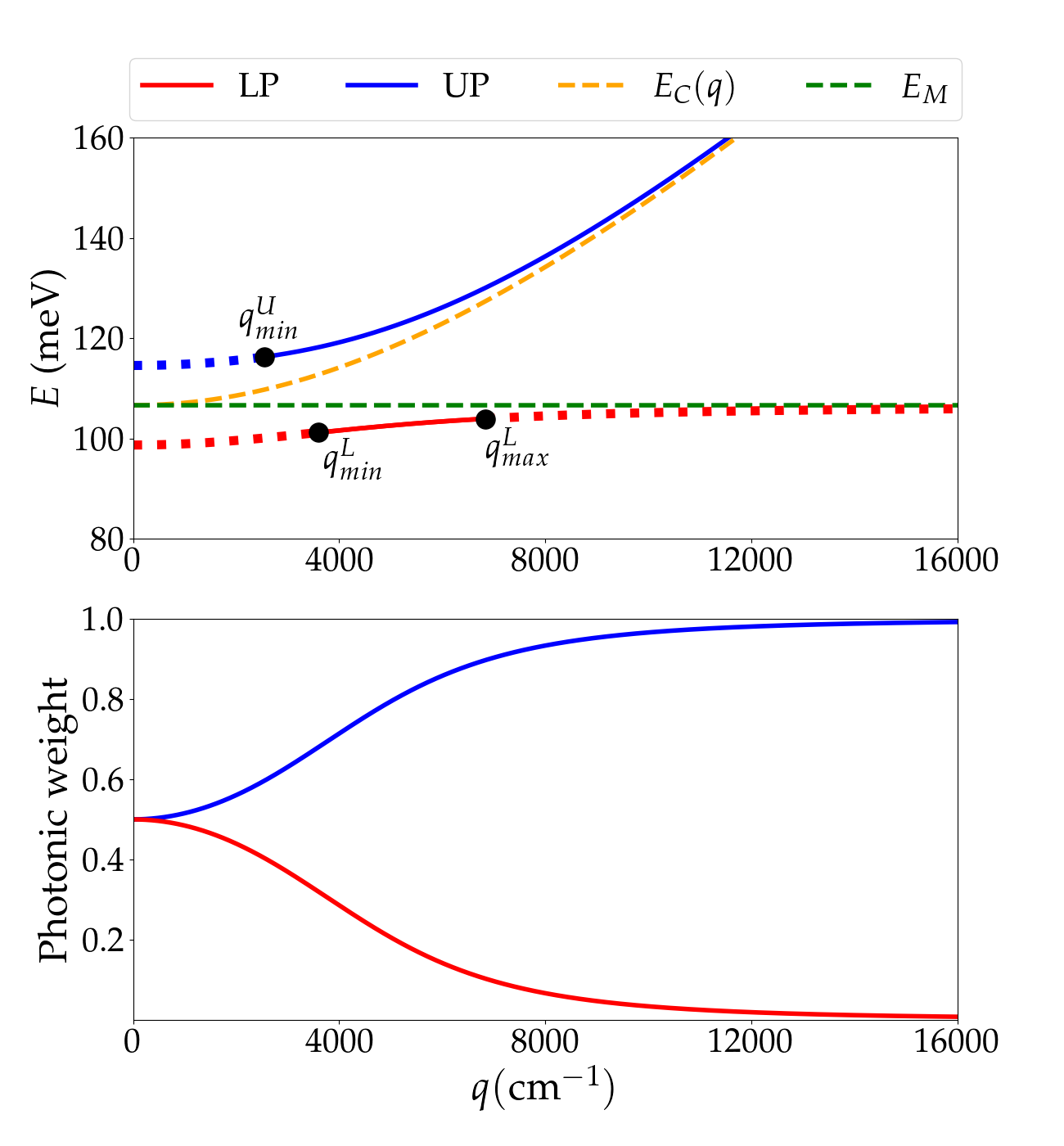}
       \caption{Top: Dispersion of plane wave-like states (continuous curves) and corresponding localization-delocalization boundaries $q_{\t{min}}$, $q_{\t{max}}$ for vibrational polaritons experimentally probed in Ref. \cite{thomas2020} (see the first row of table \ref{tab:parameters}). The low and high-energy mobility edges $q_{\t{min}}^{\t{P}}$ ($\t{P} = \t{L}$ or $\t{U}$ for LP and UP, respectively) and $q_{\t{max}}^{\t{L}}$ separate delocalized modes from the localized states satisfying the IRC with dominant inelastic scattering $2\pi l_{\t{inel}}(q) < \lambda(q)$. Bottom: Photon content of the corresponding UP and LP bands.}
    \label{fig:dispersion_curves_thomas}
\end{figure}

In order to obtain a typical transport phase diagram for a material under VSC, we first examined the system studied experimentally in Ref. \cite{thomas2020} consisting of VSC between $-$C$-$Si (PFA) with an IR microcavity at zero detuning (other relevant parameters are listed on the first row of Table \ref{tab:parameters}). Fig. \ref{fig:dispersion_curves_thomas} shows the LP and UP polariton dispersion curves for this system, along with their mobility edges $q_{\t{min}}^\t{L}$, $q_{\t{max}}^{\t{L}}$ and $q_{\t{min}}^{\t{U}}$ separating the macroscopically delocalized LP and UP modes from their localized counterparts ($q < q_{\t{min}}^{\t{L}}$ and $q > q_{\t{max}}^\t{L}$ for LP and $q < q_{\t{min}}^{\t{U}}$ for UP). In this system, inelastic scattering is stronger than elastic for all delocalized LP modes which satisfy $\lambda < 2\pi l_{\t{el}}$ and $\lambda < 2\pi l_{\t{inel}}$ simultaneously. Therefore, the  localization-delocalization boundaries, according to the IRC, are all determined by inelastic scattering. All UP modes with $q < q_{\t{min}}^{\t{U}} = 2551~\t{cm}^{-1}$ are localized (in the thermodynamic limit), but their localization length is expected to be macroscopic since their photonic content is greater than $50\%$ and $q \approx 0$. Note the absence of high-energy delocalization-localization UP boundary as follows from the fact that at large $q$ the UP is dominated by its photonic character and is only weakly affected by the disorder in the molecular system. 

Conversely, LP has both low- and high-energy mobility edges at $q_{\t{min}}^{\t{L}} = 3595~\t{cm}^{-1}$ and $q_{\t{max}}^{\t{L}} = 6843~\t{cm}^{-1}$. Both are determined by inelastic scattering since $l_{\t{inel}}^\t{L}(q) < l_{\t{res}}^{\t{L}}(q)$ and $l_{\t{fluc}}^{\t{L}}(q)$ for all relevant $q$  satisfying either $ql_{\t{inel}}^\t{L}(q) > 1$ (delocalized modes in the presence of both inelastic and elastic scattering) or $q l_{\t{el}}^{\t{L}}(q) > 1$ (delocalized modes according to the IRC only in the absence of inelastic scattering). 

\begin{figure}[h]
\includegraphics[width=\columnwidth]{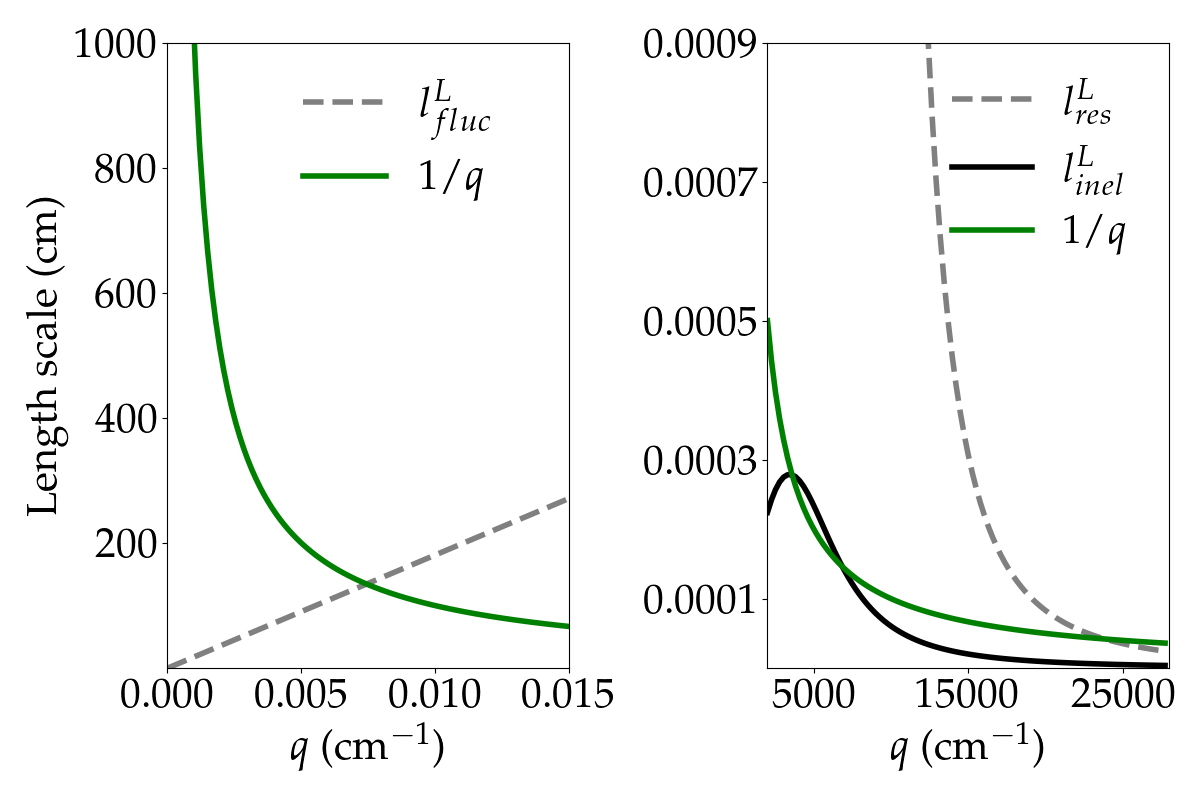}
    \caption{Comparison of mean free paths associated to scattering induced by structural fluctuations (left) $l_{\t{fluc}}^{\t{L}}(q)$,  and resonance $l_{\t{res}}^{\t{L}}(q)$, and inelastic $l_{\t{inel}}^{(L)}(q)$ (right) processes for the system experimentally studied in Ref. \cite{thomas2020} with largest Rabi splitting (see first row of Table \ref{tab:parameters}). The figures show that elastic scattering (resonance and medium fluctuation-induced) is relatively weak compared to inelastic processes [$l_\t{inel}^{\t{L}}(q) < l_{\t{res}}^{\t{L}}(q)$ and $l_\t{fluc}^{\t{L}}(q)$] for all $q$ where LP is delocalized according to the IRC.}
    \label{fig:mfp_thomas}
\end{figure}

In Fig. \ref{fig:mfp_thomas}, we provide $1/q = \lambda(q)/2\pi$ and the mean free paths associated with fluctuation, resonance, and inelastic scattering for the studied system. The dominance of decay processes induced by inelastic over resonance (elastic) scattering here observed is a consequence of the small inhomogeneous broadening ($\sigma_M \approx 1~ \t{meV}$) of the molecular system relative to the photon leakage rate ($\gamma_C \approx 3.5 ~\t{meV}$) and the molecular dephasing rate $(\gamma_M \approx 4~\t{meV})$. Note that the resonance scattering mean free path eventually becomes shorter than the inelastic mean free path at the strong localization region where $q > 2.5\times 10^4~\t{cm}^{-1}$ and both elastic and inelastic mean free paths are much smaller than the corresponding  in-plane LP wavelength $2\pi/q$.
The detuning-dependent transport phase diagram for the system investigated in this section \cite{thomas2020} is shown in Fig. \ref{fig:thomas_phasediagram3} for variable detuning $u \in [-26, 63]~\t{meV}$. Here, we find, in agreement with earlier studies \cite{litinskaya2006, litinskaya2008, ribeiro2022multimode}, that a more negatively detuned optical cavity favors delocalization by increasing the density of macroscopically delocalized LP modes. In addition, we find that for detuning $u$ greater than approximately 0.2 meV, the LP phase-breaking length $l_{\t{inel}}^{\t{LP}}(q)$ is smaller than the in-plane wavelength for all wave numbers, and no macroscopic delocalized modes are likely to exist. In any case, there remains a finite density of strongly damped propagating modes with $q_{\t{min,res}}^{\t{L}}<q < q_{\t{max,res}}^{\t{L}}$ for all $u  \in (0.2 ~\t{meV}, 63~\t{meV})$ (Fig. \ref{fig:thomas_phasediagram3} and Table \ref{tab:classification}). At detunings $u$ greater than 63 meV, only localized LP and weakly coupled molecular modes exist according to the employed classification scheme.

\begin{figure}[h] 
\includegraphics[width=\columnwidth]{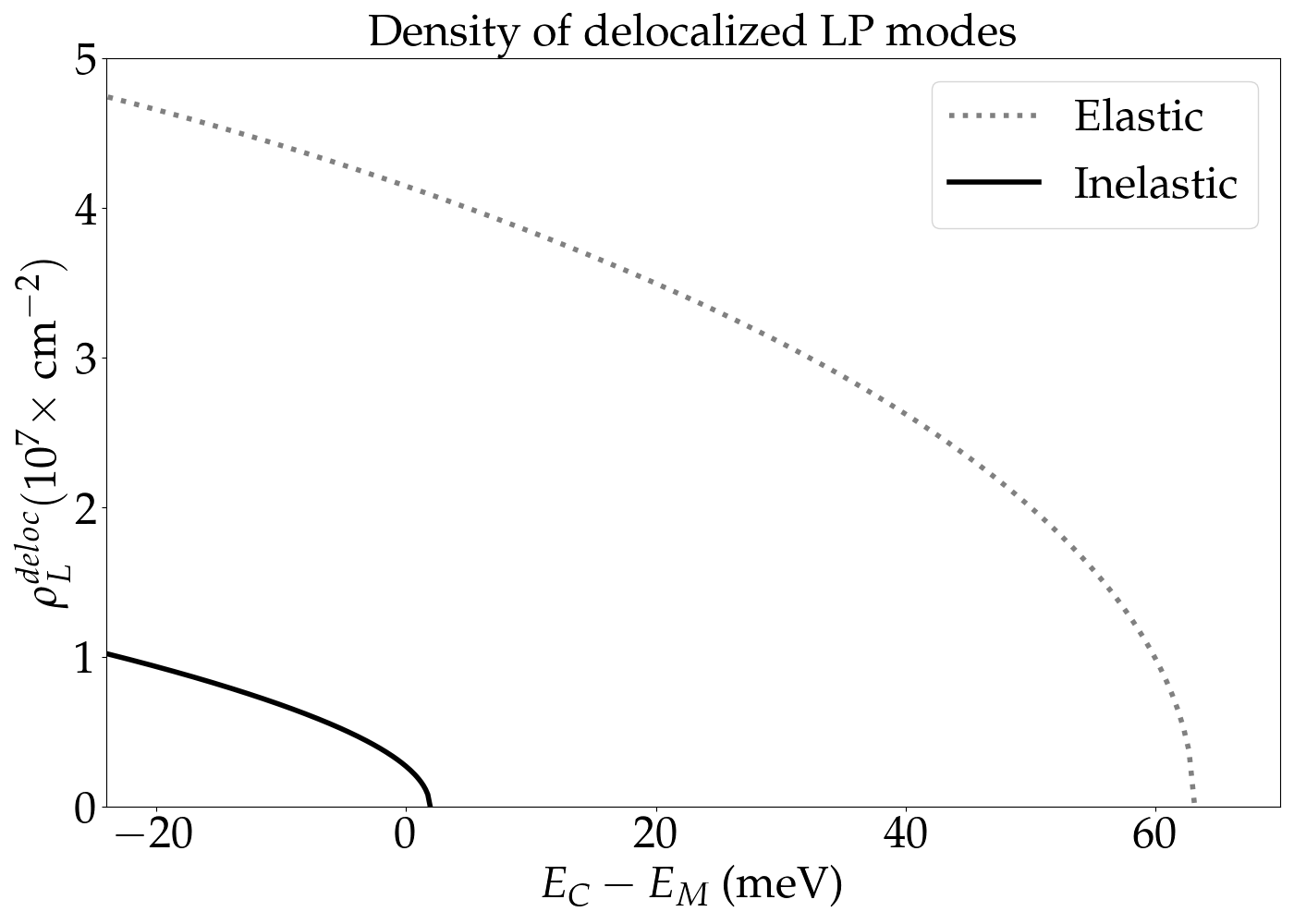}
\caption{Detuning dependence of the density of delocalized LP modes for the system in Ref. \cite{thomas2020} with strongest Rabi splitting (see the first row of Table \ref{tab:parameters}) in the presence (continuous curve) and absence of inelastic scattering (dotted curve).}    \label{fig:delocalized_modes_thomas}
\end{figure}

In Fig. \ref{fig:delocalized_modes_thomas}, we utilize the mobility edges obtained in Fig. \ref{fig:thomas_phasediagram3} to quantify the density of delocalized LP modes per unit area for variable detuning. We also report $\rho_\t{L}^{\t{deloc}}(u)$ obtained in the absence of inelastic scattering (``elastic" curve). While the number of short and long-lived delocalized LP modes per unit area increases as the detuning becomes more negative, the effect of detuning on the density of delocalized LP modes is small, especially in the presence of inelastic scattering, which drastically reduces $\rho_\t{L}^{\t{deloc}}(u)$ for $u \leq 0$.

\begin{figure}[h]
\includegraphics[width=\columnwidth]{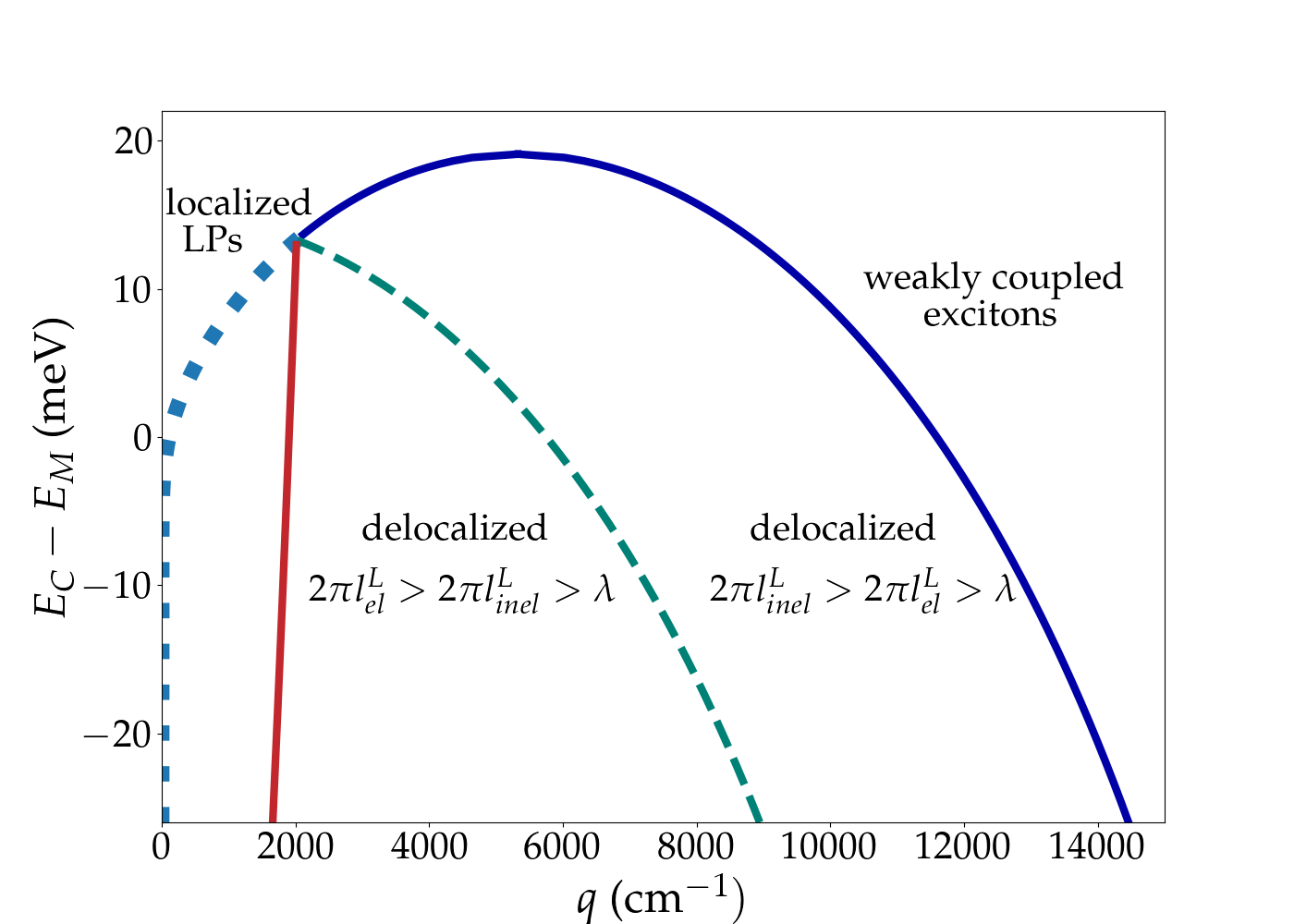}
\caption{Transport phase diagram for the inhomogeneous limit (see Table \ref{tab:parameters}) of a vibrational polariton system examined in Ref. \cite{thomas2020}. All parameters are from \cite{thomas2020} except for the variable detuning and the molecular homogeneous line width $\gamma_M$, which is set to zero in order to study the transport diagram in the limit where the molecular line shape is dominated by inhomogeneous broadening ($\sigma_M = 2.91 ~\t{meV}$ and $\gamma_M = 0$). Note that inelastic scattering remains present since polariton decay can occur via photon leakage ($\gamma_C = 3.47$ meV). The green dashed line represents the boundary between delocalized LP modes where elastic and inelastic scattering prevails. The weakening of inelastic processes relative to the system examined in Fig. \ref{fig:thomas_phasediagram3} leads to a new region ($q > q^*$, where $q^*$ is given by the green dashed line) where resonant elastic scattering dominates over inelastic, and weak wave localization and coherent backscattering phenomena are expected.}
\label{fig:thomas_phasediagram1}
\end{figure}

\begin{figure}[h]
\includegraphics[width=\columnwidth]{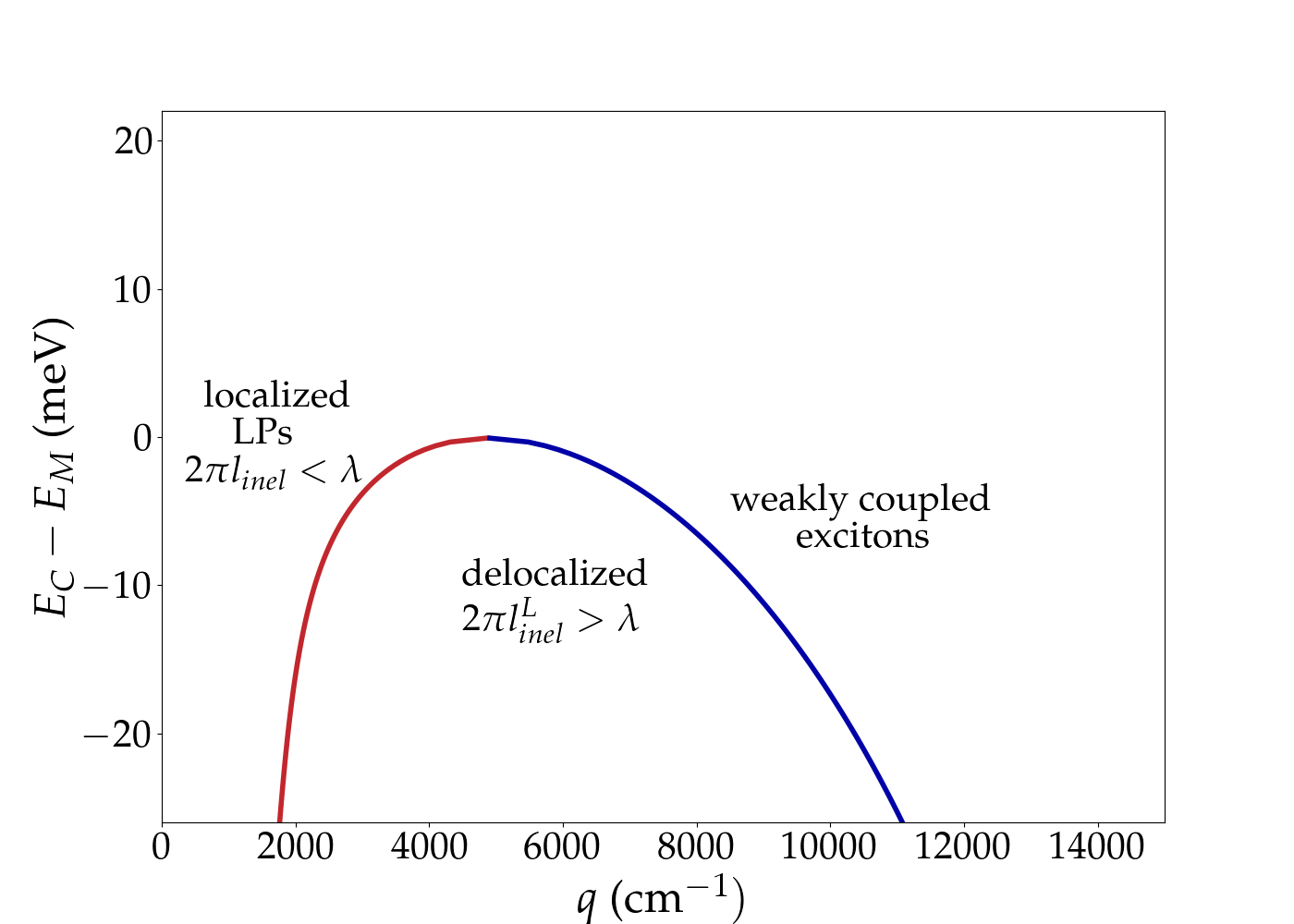}
\caption{Transport phase diagram for the homogeneous limit (see Table \ref{tab:parameters}) of the vibrational polariton system of Ref. \cite{thomas2020} with all parameters as in Figs. \ref{fig:thomas_phasediagram3} and \ref{fig:thomas_phasediagram1}, except that here $\sigma_M = 0 ~\t{meV}$ and $\gamma_M = 4.84~\t{meV}$ (obtained on the assumption that the bare molecule line width is entirely due to fast dephasing processes).}	
\label{fig:thomas_phasediagram2}
\end{figure}

To assess the impact of static and dynamical molecular disorder on detuning dependent transport phase diagrams, we present in Figs. \ref{fig:thomas_phasediagram1} and \ref{fig:thomas_phasediagram2} $q-u$ transport diagrams for the following limiting situations of molecular disorder. In Fig. \ref{fig:thomas_phasediagram1}, we take the inhomogeneous limit of the molecular line width where $\gamma_M = 0$ (and $\sigma_M$ is maximal since all the molecular line width would be due to static disorder$-$ see second row of Table \ref{tab:parameters}), whereas Fig. \ref{fig:thomas_phasediagram2} examines the vibrational polaritons emergent in the homogeneous limit of the molecular line width where $\sigma_M = 0$ (and $\gamma_M$ is maximal since all the molecular line width would be due to  homogeneous dephasing $-$ see third row of Table \ref{tab:parameters}).

In the inhomogeneous limit (Fig. \ref{fig:thomas_phasediagram1}), the low-energy mobility edge for LP transport is determined by inelastic scattering (produced in this case only by photonic leakage) for all $u$ less than approximately  14 meV, and by resonant scattering for all $u > 14$ meV. Conversely, the high-energy delocalization-localization boundary is always determined by resonance scattering, since in the limit where $\gamma_M = 0$, we find that for each detuning below the critical, there exists $q^*$ such that for $q> q^*$, elastic scattering dominates over inelastic, i.e., $l_{\t{inel}}(q)>l_{\t{el}}(q)$ (this region is to the right of the dashed line in Fig. \ref{fig:thomas_phasediagram1}). In this case, we only observe overdamped LP modes in a narrow region  $q_{\t{min,el}}^{\t{L}} < q < q_{\t{min,inel}}^{\t{L}}$ where $l_{\t{inel}}(q)< 1/q <l_{\t{el}}(q)$ and the photonic weight is sufficiently large that cavity leakage leads to efficient inelastic scattering despite the absence of molecular dephasing. Note that while $\gamma_M = 0$ is an unrealistic condition, molecular dephasing can be slowed down by lowering the temperature.  

In the inverse scenario where the molecular line shape is assumed to be in the homogeneous limit with $\sigma_M=0$ and the molecular FWHM is determined entirely by fast dephasing processes, the LP mobility edges are strictly determined from inelastic scattering, since resonance scattering is absent and fluctuation scattering is too weak (see, e.g., Fig. \ref{fig:mfp_thomas}). Fig. \ref{fig:thomas_phasediagram2} shows the  $q-u$ phase diagram obtained under these conditions. Notably, the critical detuning (close to zero meV) is drastically smaller relative to that in the inhomogeneous limit where $\sigma_M$ is large and $\gamma_M$ vanishes (Fig. \ref{fig:thomas_phasediagram1}), but only slightly smaller relative to the experimental system which operates near its critical detuning (Fig. \ref{fig:thomas_phasediagram3}). Overall, a much richer diversity of polariton transport phenomena is expected for a system with weak inelastic and strong elastic scattering (Fig. \ref{fig:thomas_phasediagram1}) in comparison to the opposite case with strong inelastic and weak elastic scattering processes (Fig. \ref{fig:thomas_phasediagram2}). 

\subsection{Disorder - wave number transport phase diagrams at zero detuning}\label{ssec:disorder}
In this subsection, we examine the impacts of static and dynamic disorder on the localization of molecular vibrational LP modes at fixed zero detuning. In analogy with studies of metal-insulator Anderson transitions \cite{sheng2006introduction, evers2008}, we present mobility edges, identify regions of the quasiparticle spectrum where (coherent) diffusion is absent and estimate critical values of disorder above which all polaritons are strongly localized according to the IRC.

\begin{figure}
\includegraphics[width=\columnwidth]{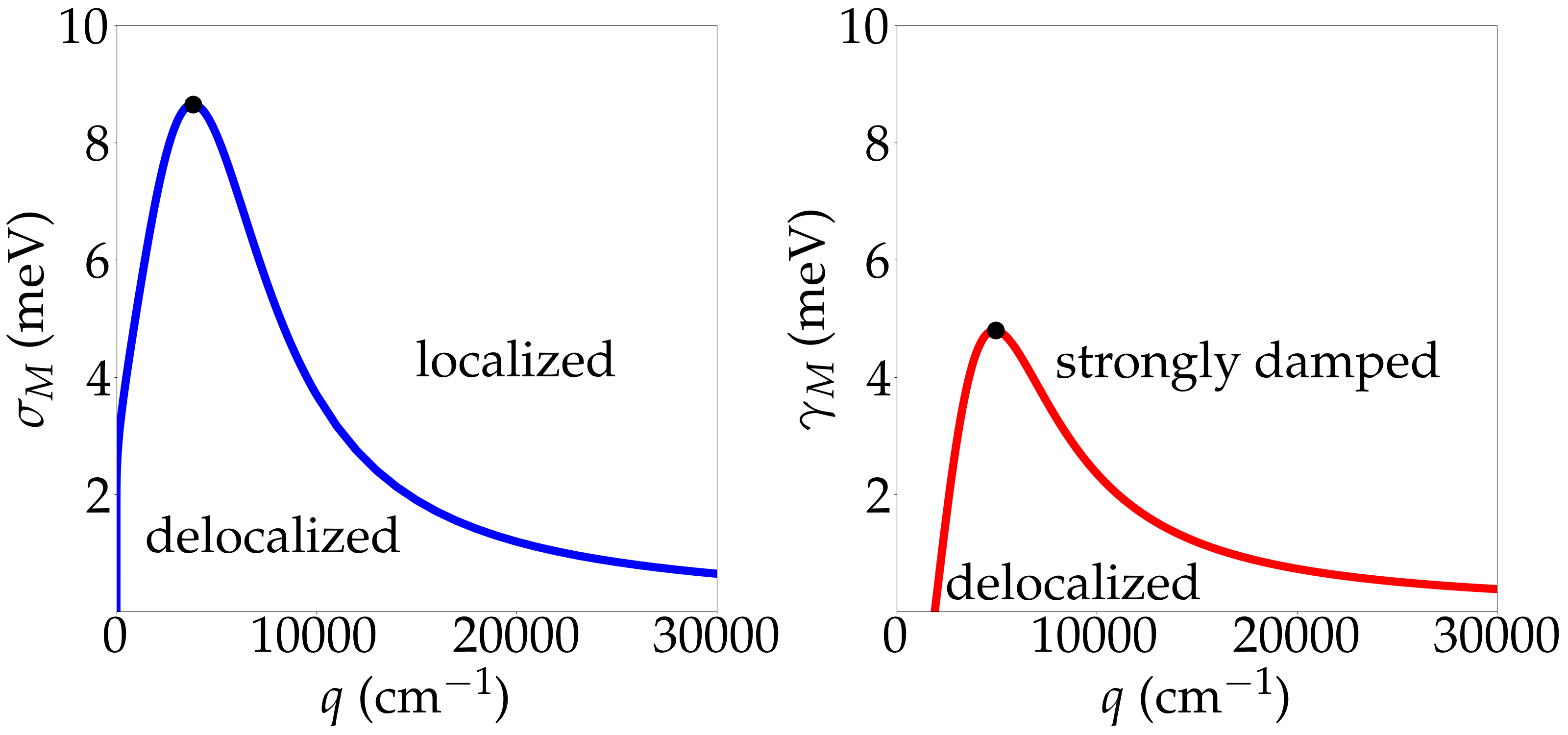}
\caption{Left: $(\sigma_M, q)$ phase diagram for system in Ref. \cite{thomas2020} in the inhomogeneous limit of the molecular system (with $\gamma_M = 0$, see Table \ref{tab:parameters}). Right: $(\gamma_M, q)$ phase diagram for the system in Ref. \cite{thomas2020} in the homogeneous limit of the molecular system (with $\sigma_M = 0$, see Table \ref{tab:parameters}).}	
\label{fig:thomas_phasediagrams_disordergamma}
\end{figure}

Figure \ref{fig:thomas_phasediagrams_disordergamma} shows the $ q - \sigma_M$ and $q - \gamma_M$ transport phase diagrams and corresponding critical values for the system with the largest collective light-matter interaction strength ($\hbar\Omega_R$ = 15.9 meV) among those examined by Thomas et al. in Ref. \cite{thomas2020}.  We find the critical energetic disorder beyond which \textit{all} LP modes are localized according to the IRC is given by $\sigma_{M}^{\t{c}} \approx 0.55 \times \hbar \Omega_R$, whereas the critical molecular dephasing rate $\gamma_M^{c}$ is characterized by the weaker requirement $\gamma_M^{c} \approx 0.3\times \hbar \Omega_R$. The condition on $\gamma_M$ for strong damping is significantly milder than that for strong Anderson localization of the analogous closed system, including only static energetic disorder. However, it is important to note that both $\gamma_M^{\t{c}}$ and $\sigma_M^{\t{c}}$ increase nontrivially with increasing $\Omega_R$ and decreasing detuning. 

Notably, while $\gamma_M$ and $\sigma_M$ are related to the line widths of the bare molecule transition, they contribute to inelastic and elastic scattering in qualitatively distinct ways. Specifically, $\gamma_M$ is directly proportional to the molecular contribution to the LP inelastic scattering rate (Eq. \ref{eq:gamma_inel}), whereas the same is not necessarily true for $\sigma_M$, as this quantity controls the magnitude of the LP resonance scattering rate by modulating the bare molecule density of states at the LP energy (see Eq. \ref{eq:res_scatter}). A final caveat about a direct comparison between $\sigma_M^c$ and $\gamma_M^c$ is that, as is well known for the Anderson model, critical points for the wave function delocalization-localization transition depend on the disorder probability distribution function, and thus, it is possible that other distributions such as the uniform may give weaker requirements for a localization-delocalization polariton transition.

 The diagrams presented in Fig. \ref{fig:thomas_phasediagrams_disordergamma} are similar to those obtained in studies of Anderson localization. This suggests that $\sigma_{M}^{\t{c}}$ is the critical point of a continuous phase transition.
 \subsection{Effects of light-matter interaction strength on LP delocalization at zero detuning}\label{ssec:rabi}
\begin{figure}[h]
\includegraphics[width=\columnwidth]{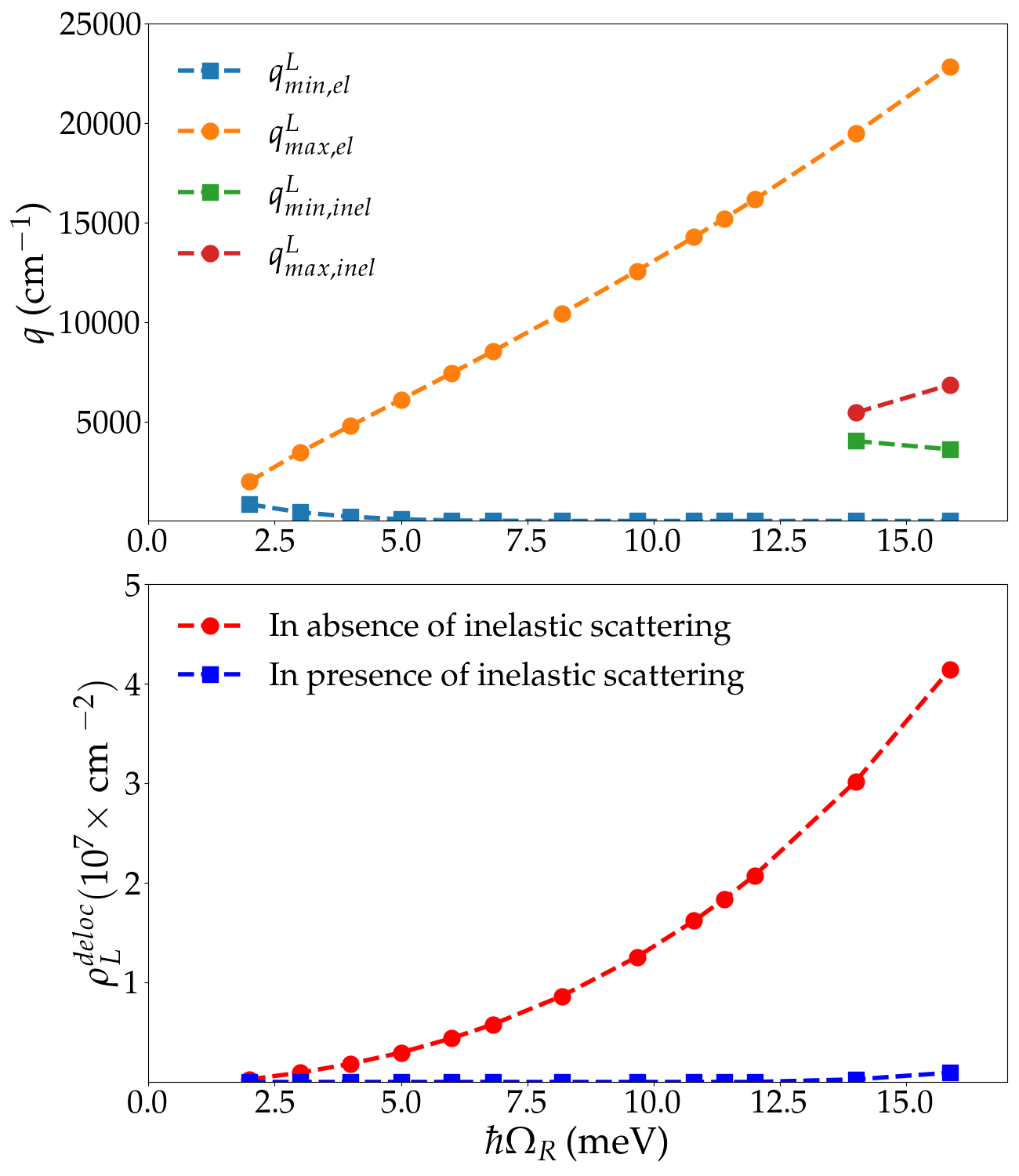}
     \caption{Top: Zero-detuning low and high-energy mobility edges for the delocalization-localization transition of LP according to the IRC as a function of collective light-matter interaction strengths probed in Ref. \cite{thomas2020}. Bottom: Density of macroscopically delocalized LP modes for the same systems as in the top figure.}
\label{fig:thomas_boundaries_fractions}
\end{figure}

To identify the effect of variable light-matter interaction strength on the localization-delocalization boundaries of vibrational polaritons, we examine the systems experimentally probed in Ref. \cite{thomas2020} with $\hbar\Omega_R(\t{in meV}) = 15.9, 12.0, 11.4, 9.7, 8.2,$ and 6.8, and intermolecular distance $a = 0.69~\t{nm}$ for $\hbar\Omega_R = 15.9~\t{meV}$ (the values of $a$ for the other $\Omega_R$ are approximated from $\Omega_R \propto \sqrt{N_M/V} \approx a^{-3/2}$). Disorder and microcavity parameters are equal for these systems (given in the first row of Table \ref{tab:parameters}). The behavior of the high-energy LP and UP mobility edges and the density of delocalized LP modes as a function of $\Omega_R$ are presented in Fig. \ref{fig:thomas_boundaries_fractions} (we also computed localization-delocalization boundaries for additional interaction strengths not examined experimentally in Ref. \cite{thomas2020}).

Figure \ref{fig:thomas_boundaries_fractions} shows that in the presence of inelastic scattering, the Rabi splitting has almost no effect on the density of delocalized LP modes. Furthermore, the IRC (Eq. \ref{eq:irc}) implies macroscopically delocalized LP modes only exist at the largest probed collective light-matter interaction strengths, namely 14 and 15.9 meV. 

The situation is significantly more interesting when inelastic processes are ignored. In particular, while the low-energy LP localization-delocalization boundary set by elastic (resonance) scattering is essentially independent of $\Omega_R$, Fig. \ref{fig:thomas_boundaries_fractions} shows that the high-energy mobility edge $q_{\t{max, el}}^{\t{L}}$ and the density of delocalized LP modes $\rho_{\t{L}}^{\t{deloc}}$ increases exponentially with $\Omega_R$. This behavior is ascribed to $q_{\t{min}}^{\t{L}}$ being close to zero for all $\Omega_R$, while the $q_{\t{max},\t{el}}^{\t{L}}$ are determined by resonance scattering that gets largely suppressed at a fixed LP wave number $q$ with increasing $\Omega_R$. The substantial effect of $\Omega_R$ on the resonance scattering mean free path (for modes with moderate to large $q$) occurs as a result of larger $\Omega_R$ leading to greater energetic separation between the LP and the center of the molecular reservoir density of vibrational modes. 

Our finding that the density of (strongly damped) propagating LP modes increases with $\Omega_R$ resembles the key observation of Ref. \cite{thomas2020} that the cavity effect on reactivity also increases with $\Omega_R$. 

To conclude, we contrast the detuning and Rabi splitting dependence of $\rho^{\t{deloc}}_\t{L}$ presented in Figs. \ref{fig:delocalized_modes_thomas} and \ref{fig:thomas_boundaries_fractions} (bottom), respectively. These figures show that increasing light-matter interaction strength is much more effective at inducing the formation of propagating LP modes than reducing the light-matter detuning. This issue is relevant because experimental work suggests that zero-detuning is optimal for maximizing polariton effects on chemistry. While our findings suggest otherwise that larger values of $\rho_\t{L}^{\t{deloc}}$ occur at negative detuning, the difference with respect to zero-detuning (``on-resonance" case) is small. 

\subsection{Comparison between electronic exciton and vibrational polaritons}\label{ssec:compare}

In this final subsection, we make a comparison between electronic exciton and vibrational polaritons. Table 1 presents the main parameters required to obtain the polariton dispersion and mobility edges from the application of the IRC to the organic microcavity system of Ref. \cite{lerario2017}. Our choice to examine this system was motivated by the large Rabi splitting and the superfluid phase transition achieved under intense pumping. 

\begin{figure}[h]
\includegraphics[width=\columnwidth]{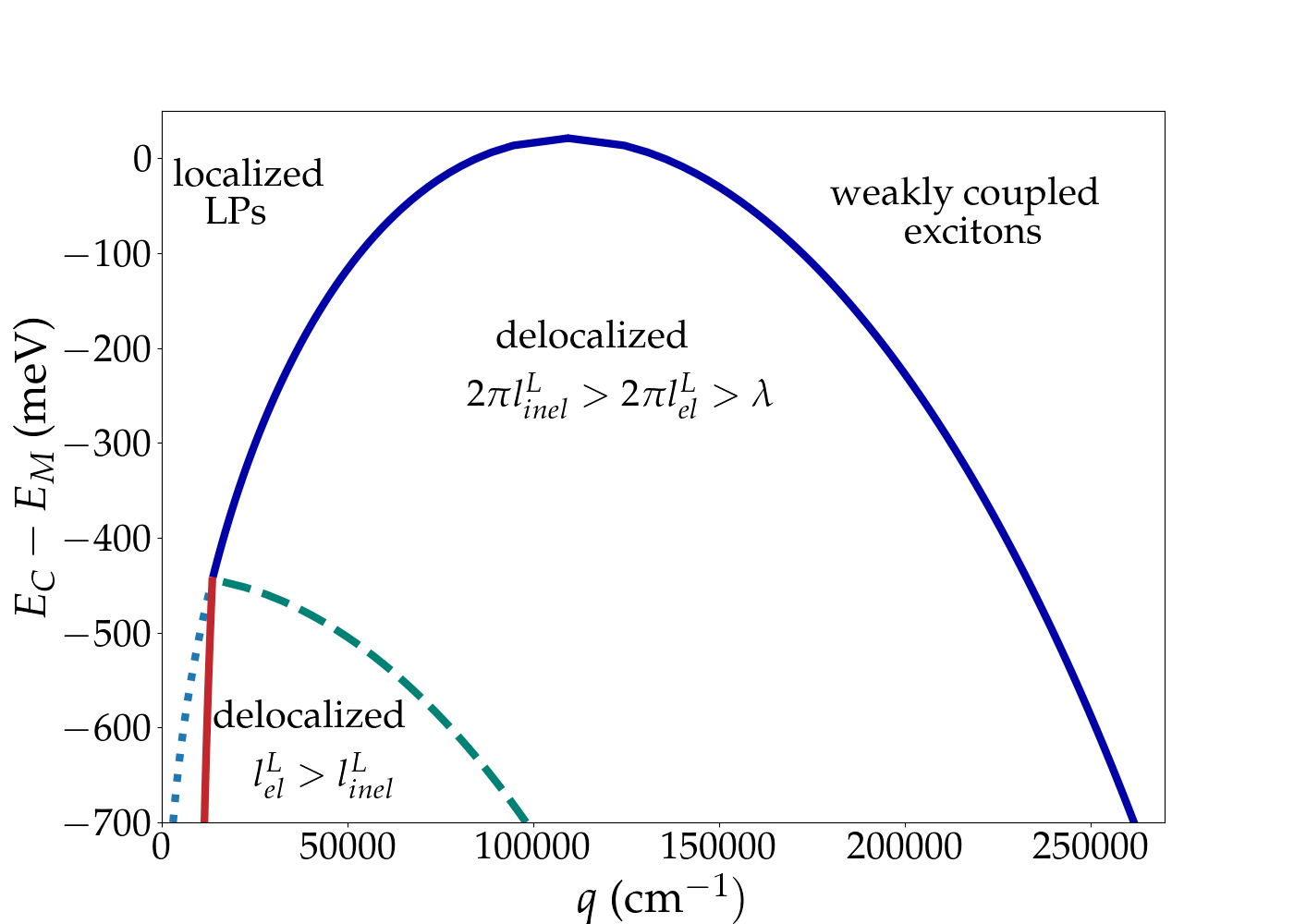}
	\caption{LP transport phase diagram for the electronic TDAF exciton-polariton system in Ref. \cite{lerario2017} (see Table \ref{tab:parameters}).}\label{fig:organic_comparison}
\end{figure}

\begin{figure}[h]
   \includegraphics[width=\columnwidth]{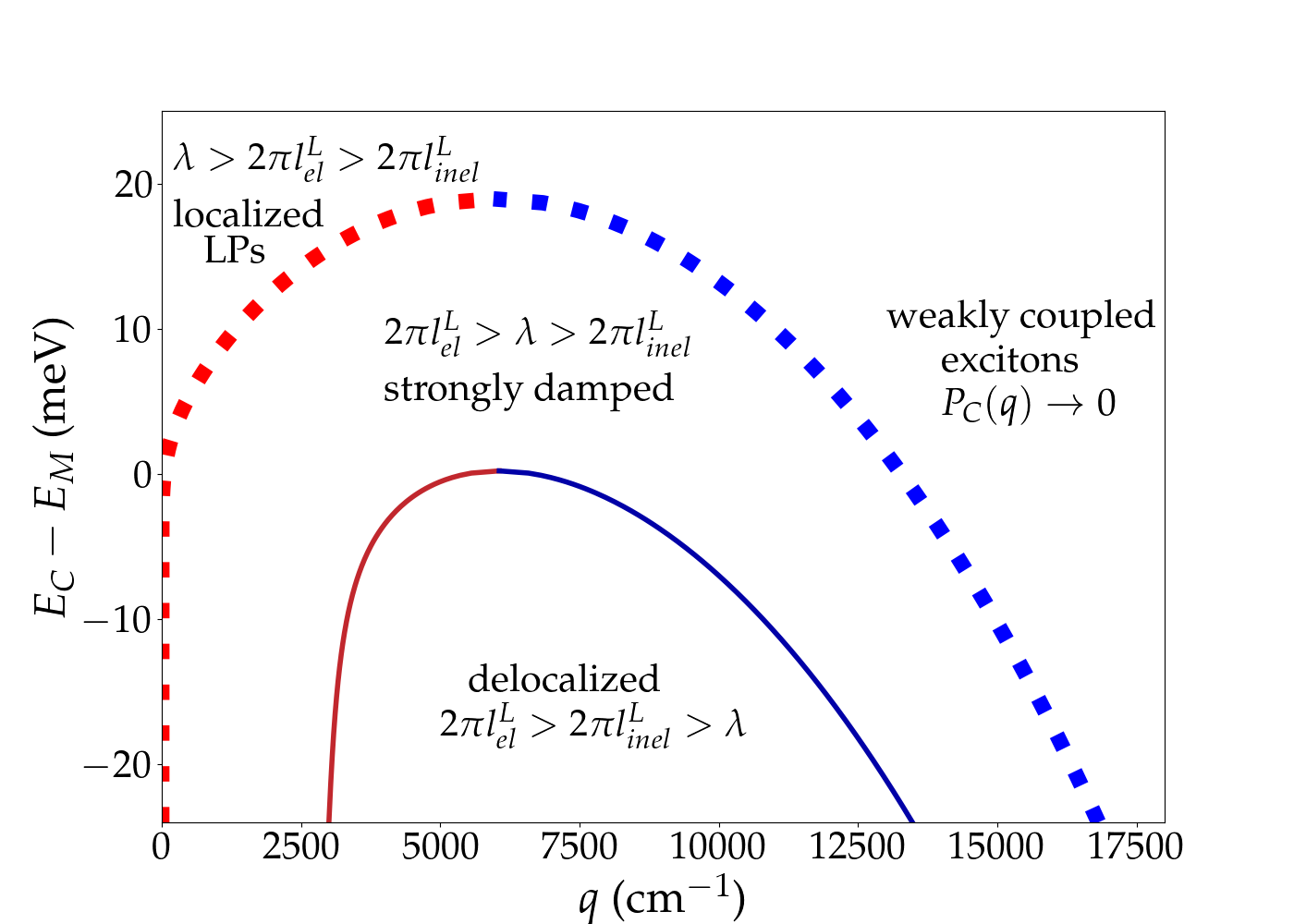} 
   \caption{Transport phase diagram for LP modes emergent from vibrational strong coupling between $-\t{NCO}$ normal-modes and an infrared microcavity \cite{simpkins2015} with parameters listed in Table \ref{tab:parameters}}\label{fig:phase_diagram_simpkins}
\end{figure}

To provide evidence of the generality of our prior analysis of vibrational polariton phases, we compare the obtained organic electronic exciton-polariton transport diagram in Fig. \ref{fig:organic_comparison} with the detuning dependent vibrational polariton transport diagram (Fig. \ref{fig:phase_diagram_simpkins}) obtained for the LP modes arising from the strong coupling of $-$NCO stretch modes with an IR microcavity experimentally studied by Simpkins and coworkers \cite{simpkins2015} (the experimental parameters describing this light-matter system are listed in Table \ref{tab:parameters}).

A key feature of the presented electronic (organic) exciton-polariton transport phase diagram is the dramatically broader interval of wave numbers supporting macroscopic delocalization in comparison to the vibrational polariton system of Fig. \ref{fig:phase_diagram_simpkins}. At the experimentally investigated detunings ($u = -455~\t{meV}$ for the electronic (organic) exciton-polariton system and $u = -12~\t{meV}$ for the vibropolariton system), the prediction of the IRC for the density of delocalized LP modes of the electronic exciton-polariton is larger by three orders of magnitude relative to the IR system.

It is notable that the electronic exciton-polaritons show a much smaller density of overdamped modes in comparison to the IR polaritons. This behavior is due to the greater energetic disorder of the electronic  excitons, which leads to resonance scattering dominance over inelastic processes even at relatively small $q$ (Fig. \ref{fig:organic_comparison}). A similar feature only arose for the vibrational polaritons studied in this paper in the inhomogeneous limit examined in Fig. \ref{fig:thomas_phasediagram1}, when we analyzed the excitations arising in the limit where the entire line width of the bare molecule ensemble is due to static inhomogeneity of the vibrational transition energies and $\gamma_M$ was assumed to vanish. Roughly, we observe, in general, that transport phase diagrams with significant regions of the wave number space including delocalized LP modes with $l_{\t{inel}} > l_{\t{el}}$ (e.g., Figs. \ref{fig:organic_comparison} and \ref{fig:thomas_phasediagram1}) occur only when  $\sigma_M / \gamma_{M} \gg 1$. \\

The observed suppressed scattering and large density of delocalized electronic exciton LP modes in Fig. \ref{fig:organic_comparison} are likely key features enabling the observation of quasi-long-range off-diagonal order and superfluidity in this system \cite{lerario2017}. Typical vibrational polariton systems (e.g., Figs. \ref{fig:thomas_phasediagram3} and \ref{fig:phase_diagram_simpkins}) will likely exhibit challenges reaching a similar quasi-Bose Einstein Condensate (quasi-BEC) regime in large part due to the significantly stronger inelastic scattering and reduced density of vibrational LP modes which lead to generic vibrational polariton phase diagrams resembling Figs. \ref{fig:thomas_phasediagram3} and \ref{fig:phase_diagram_simpkins} which have a small density of delocalized LP modes according to the IRC.
\section{Conclusions}\label{sec:conclusions}
We applied the Ioffe-Regel criterion to construct transport phase diagrams and identify the main features of transport phenomena of vibrational polaritons emergent from collective strong interactions of molecular vibrational modes and planar IR microcavities. We unraveled optimal conditions of negative detuning, low disorder, and large Rabi splitting for the maximization of delocalization in vibrational polaritons. 

Lower polariton modes were classified as (macroscopically propagating) delocalized, (propagating) strongly damped, localized LPs, and as localized weakly coupled excitons. Each of these categories corresponds to a qualitatively distinct regime of polariton delocalization and transport that was characterized in various experimentally studied systems. 

We found that despite the reduced density of delocalized vibrational LP modes relative to the weakly coupled, a finite and significant density of macroscopic delocalized vibrational LPs persists in general, even at zero detuning. Macroscopic delocalization was found to be maintained under typical vibrational strong coupling conditions even for polariton states with nearly insignificant photon content (e.g., less than ten percent). Our results provide significant new insight into transport phenomena under VSC and are expected to be particularly relevant in informing future investigations of polariton-assisted processes that may largely benefit from wave function delocalization. In addition, while we make no statements about the properties of the strongly localized molecular reservoir modes (``weakly coupled excitons"), we expect that polariton-induced perturbations of local dynamical processes (e.g., reactivity and vibrational energy relaxation) involving dark modes will be greater under conditions where a larger number of polaritons are macroscopically delocalized and the vibrational polariton high-energy mobility edges occur at higher $q$. Therefore, our analysis of polariton transport phenomena is expected to be relevant in the future rationalization of polariton chemistry trends and for the establishment of optimal conditions for the control of chemical kinetics with photonic materials.

Finally, our work has provided a new perspective on challenges expected in the way of reaching quasi-BEC and superfluid behavior in the VSC regime. By comparing the delocalization in electronic (organic) exciton and vibrational polariton systems, we found the former tend to have a density of macroscopic delocalized modes several orders of magnitude larger than the latter. This suggests that the pumping threshold for the achievement of an LP nonequilibrium BEC phase under VSC will be much larger relative to that required for electronic (organic) exciton condensation since the condensate is formed from the establishment of long-range off-diagonal order, which is facilitated by wave function delocalization. \\

\begin{center} 
\textbf{ACKNOWLEDGMENTS}\\
\end{center}

We acknowledge generous start-up funds from the Emory University Department of Chemistry.\\

\begin{center} 
\textbf{DATA AVAILABILITY}\\
\end{center}

The data that supports the findings of this study are available from the corresponding author upon reasonable request.\\

\begin{center}
\textbf{Appendix A: Rate of inelastic scattering}\\
\end{center}
\tb{In this Appendix, we provide a more detailed discussion of the inelastic scattering processes considered in the main manuscript and a derivation of Eq. \ref{eq:gamma_inel}. We assume the inelastic scattering events are uncorrelated and may be ascribed to three qualitatively distinct processes corresponding to cavity leakage, molecular population relaxation, and molecular dephasing. For concreteness, the light-matter system is supposed to be in the vibrational strong coupling regime. We begin with the general Hamiltonian below, including intermolecular interactions and system-bath coupling,}
\begin{align}
H = H_\t{L} + H_\t{M} +H_{\t{LM}} + H_{\t{B}} + V_{\t{MM}}+ V_{\t{SB}}, \label{eq:app_ham}
\end{align}
\tb{where $H_\t{L}$, $H_\t{M}$, and $H_{\t{LM}}$ are the same as in the main text, $V_{\t{MM}}$ contains quadratic nearest-neighbor electrostatic dipole-dipole interactions (see Appendix B), $H_{\t{B}} = H_{\t{B}}^C + H_{\t{B}}^M$, where $H_{\t{B}}^C$ and $H_{\t{B}}^M$ are Hamiltonians for the bath weakly coupled to the cavity and molecular subsystems, and $V_{\t{SB}} = V_{\t{SB}}^{C} + V_{\t{SB}}^{M}$ represents all light and matter interactions with their corresponding baths ($V_{\t{SB}}^{C}$ and $V_{\t{SB}}^{M}$, respectively). Each cavity mode with in-plane momentum $\mathbf{q}$ and polarization $\lambda$ (we omit the index $m$ since we consider a single cavity band to be relevant) is assumed to interact bilinearly with an independent set of free space harmonic modes with creation and annihilation operators $c_{\mbf{q}\lambda\omega}^{s,\dagger}$ and $c_{\mbf{q}\lambda\omega}^s$, with $\omega >0$ and $s= L~\t{or}~R$ corresponding to free space modes to the left and right of the cavity, respectively \cite{gardiner1985}}
\begin{align}
 V_{\t{SB}}^{C} =  \hbar\sqrt{\frac{\gamma_C}{4\pi}}\sum_{s = L,R}\sum_{\mbf{q}\lambda}\int_0^{\infty}\mrm{d}\omega~\l[a_{\mbf{q}\lambda}^\dagger c_{\mbf{q}\lambda\omega}^s   +c_{\mbf{q}\lambda\omega}^{s,\dagger} a_{\mbf{q}\lambda}\r],
\end{align}
\tb{where we assume without loss of generality, the cavity decay rate $\gamma_C$ is frequency and polarization independent.}

\par \tb{The molecular system-bath interactions are described as follows. Each molecule $i$ interacts independently with a set of low and high-frequency bath modes $\{\alpha\}$, with creation and annihilation operators $b_{i\alpha}^\dagger, b_{i\alpha}$ and $B_{i\beta}^\dagger$ and $B_{i\beta}$, respectively. The low-frequency bath modes correspond to, e.g., librations with frequencies that are much lower (and thus are highly off-resonant) than $E_M/\hbar$. These modes form a quasicontinuum that induces dephasing of the molecular polarization and typically provides the dominant contribution to the molecular absorption lineshape in free space \cite{kenkre1994theory, mukamel1999principles}. In solution, solute-solvent interactions are stronger than solute-solute, and the bare molecule homogeneous broadening linewidth $\gamma_M = 1/T_2$ is often entirely determined by the interaction of the low-frequency bath modes with the matter polarization. High-frequency environment modes have a much lower density of states than low-frequency, but the former are nevertheless involved with vibrational population relaxation via excitation energy transfer involving one or multiple vibrational quanta of the system. Nevertheless, vibrational population relaxation, in general, occurs on a much longer timescale ($T_1 \gg T_2$) than dephasing induced by low-frequency modes \cite{kenkre1994theory, tokmakoff1994phonon}. For this reason, we will ignore the contribution of population relaxation to the polaritonic inelastic scattering. This suggests the following well-known representation of the matter-bath interactions \cite{mukamel1999principles, may2008charge}}
\begin{align}
V_{\t{SB}}^{M} = \sum_{i=1}^{N_M} \sum_{\alpha}\hbar \omega_\alpha \lambda_\alpha \sigma_i^+ \sigma_i^- \l(b_{i\alpha} + b_{i\alpha}^{\dagger}\r). 
\end{align}
\tb{We reiterate our assumption that each molecule has an independent set of bath modes and note that del Pino et al. \cite{pino2015} have shown that this scenario (compared to those where the bath modes interact simultaneously with multiple strongly coupled molecules) provides the fastest transition rates for polariton $\rightarrow$ weakly coupled mode in the absence of strong direct forces between the molecules involved in strong coupling (e.g., when $V_{\t{MM}}$ has negligible effects on polariton dynamics).}

\par \tb{In the mean-field limit \cite{litinskaya2006}, the normal-modes of $H_\t{M} +H_\t{LM} + V_{\t{MM}}$ consist of degenerate TE and TM polaritons and longitudinal matter polarization modes (``LO modes") which are decoupled from light \cite{vinogradov1992vibrational, kittel2005introduction}. We will assume the LO modes are dispersionless and the energy shift induced by $V_{\t{MM}}$ relative to $E_M$ is negligible compared to system-bath interactions. These assumptions are typical for effective Hamiltonians representing solute dynamics.}

\par \tb{Dynamical fluctuations of the intermolecular distance, orientation, and molecular excitation energies also enable inelastic scattering, i.e., polariton decay into weakly coupled modes\cite{li2022polariton}, corresponding in our model to strongly localized linear combinations of longitudinal modes and polaritons with nearly negligible photon content (e.g., less than $0.1\%$). In this Appendix, we focus on the inelastic scattering induced by intramolecular system-bath coupling and cavity decay. Polariton decay into weakly coupled modes via fluctuations of the direct intermolecular interaction strength is discussed in Appendix B.}

\tb{We assume the initial state of the inelastic scattering process is a direct product of a polariton pure state and the bath thermal density operator and compute the rate of polariton decay using Fermi's golden rule. This requires the evaluation of} 
\begin{align}
 & R_{FI}^{\t{SB},M} \equiv \sum_{n_{bI}, n_{bF}} p_{n_{bI}} |\braket{F,n_{bF}| V_{\t{SB}}^{M}|I,n_{bI}}|^2,\\
 &  R_{FI}^{\t{SB},C} \equiv \sum_{n_{bI}, n_{bF}} p_{n_{bI}} |\braket{F,n_{bF}| V_{\t{SB}}^{C}|I,n_{bI}}|^2,
\end{align}
\tb{where $I$ is a polariton state, $F$ is the final state corresponding either to a weakly coupled molecular mode or to the cavity-matter global ground state, $n_{bI}$ and $n_{bF}$ are initial and final bath Fock states, and $p_{n_{bI}}$ is the probability for a particular initial state of the light-matter bath. The terms $R_{FI}^{\t{SB}, M}$ with $I$ corresponding to a lower polariton mode $P(\mbf{q})$ and $F$ corresponding to  weakly coupled molecular modes $D_\mu$ with $\mu = 1,2,..., N_M-N_\t{P}$ (where $N_\t{P}$ is the number of lower polariton modes as defined above, and the set $\{\ket{D_\mu}\}$ contains LP modes with very small photonic content, such that they may be viewed as weakly-coupled excitons according to Table \ref{tab:classification}) with $E_{D_\mu} > E_{P(\mbf{q})}$ can be readily computed in the mean-field limit leading to the following rate of LP inelastic decay into the dense manifold of weakly coupled states}
\begin{widetext}
\begin{align}
    k_{D\leftarrow P(\mbf{q})}^{\t{SB},M} & = \frac{2\pi}{\hbar} \sum_\mu 
     \sum_{mn} c_{mP(\mbf{q})}^* c_{nP(\mbf{q})} c_{m D_\mu} c_{n D_\mu}^*\sum_\alpha (\hbar\omega_\alpha \lambda_\alpha)^2 \braket{n_\alpha}\delta\l[E_\alpha-(E_{D_\mu}-E_{P(\mbf{q})})\r] \nonum 
     & \approx \frac{2\pi}{\hbar} \sum_\mu P_{M}^P(\mbf{q}) P_{M}^{D_\mu}\sum_{mn} \frac{\delta_{mn}}{N_M^2}\sum_\alpha (\hbar\omega_\alpha \lambda_\alpha)^2 \braket{n_\alpha} \delta\l[E_\alpha-(E_{D_\mu}-E_{P(\mbf{q})}))\r]\nonum 
     & \approx \frac{2\pi}{\hbar} P_{M}^P(\mbf{q}) \frac{N_M-N_{\t{P}}}{N_M}  \sum_\alpha (\hbar\omega_\alpha \lambda_\alpha)^2 \braket{n_\alpha} \delta\l[E_\alpha-(E_{M}-E_{P(\mbf{q})})\r] \nonum 
     & \approx \frac{1}{\hbar} P_{M}^P(\mbf{q}) J(E_{M}-E_{P(\mbf{q})}) \braket{n(E_{M}-E_{P(\mbf{q})})}.
\end{align} \end{widetext}
\tb{where $\braket{n_\alpha}$ is the mean occupation number of the molecular bath mode $\alpha$ and $J(\omega) = \sum_{\alpha} (\hbar\omega_\alpha \lambda_\alpha)^2 \delta(E-E_\alpha)$ is the bath spectral density. Using the high-temperature limit of $\braket{n(\omega)} = k_\t{B}T/\hbar\omega$ (note that for vibrational polaritons, $E_{M}-E_{P(\mbf{q})}$ is almost always much smaller than room temperature) and the Ohmic spectral density $J(\omega) = \eta \omega e^{-\omega^2/\omega^2_{\t{cut}}}$ \cite{weiss2012quantum} with $\omega_{\t{cut}} \gg \omega_{M} - \omega_{P(\mbf{q})}$ and $\eta > 0$, it follows that}
\begin{align}
   k_{D\leftarrow P(\mbf{q})}^{\t{SB},M} & \approx \frac{\eta  k_\t{B}T}{\hbar} P_{M}^P(\mbf{q}) \nonum 
   & \approx \gamma_M P_{M}^P(\mbf{q}), \label{eq:matter_decay}
\end{align}
\tb{where we used  $\eta  k_\t{B}T/\hbar = \gamma_M$, thus obtaining the matter contribution to Eq. \ref{eq:gamma_inel} of the main text. This result agrees with that obtained by del Pino et al. \cite{pino2015}, who examined the effects of system-bath interactions on vibrational strong coupling between a cavity photon mode and a molecular ensemble \cite{pino2015}.} 
\par \tb{The computation of the photonic contribution to the inelastic scattering rate via Fermi's Golden rule follows the same ideas as above, except the system final state $F$ is the light-matter ground state, and the polariton decay process is induced by the photonic bath}
\begin{align}
     k_{G \leftarrow P(\mbf{q})}^{\t{SB},C} = P_C^P(\mbf{q}) \gamma_C . \label{eq:cav_decay}
\end{align}

\par \tb{In the absence of sufficiently strong interactions between the molecular modes contributing to polariton formation, the total rate of polariton inelastic scattering  (Eq. \ref{eq:gamma_inel}) follows from Eqs. \ref{eq:matter_decay} and \ref{eq:cav_decay}. Molecular population relaxation was also ignored since matter dephasing induced by coupling to a low-frequency bath tends to be much more effective at inducing polariton transitions than vibrational population relaxation processes, which typically take much longer than polarization dephasing \cite{pino2015}.}

\begin{center}
\textbf{Appendix B: Polariton inelastic scattering via intermolecular dephasing pathway}
\end{center}
\tb{Fluctuations in the intermolecular interactions between molecular modes contributing to collective strong light-matter coupling may also lead to polariton dephasing \cite{li2022polariton}. Assuming this mechanism is incoherent, mediated by dipole-dipole interactions, and the molecular dipole vectors are independently distributed over the unit sphere, the polariton inelastic scattering induced by this process has rate \cite{li2022polariton}}
\begin{align}
    k_{D\leftarrow P(\mbf{q})}^{\t{dd},M} = 2\pi \Delta^2 P_M^P(\mbf{q}) S_{P(\mbf{q})}, \label{eq:kdpqdd}
\end{align} 
\tb{where $S_{P(\mbf{q})}$ is the spectral overlap integral between the polariton $P(\mbf{q})$ and the weakly coupled mode reservoir}
\begin{align}
     S_{P(\mbf{q})} = \int_0^{\infty} \mrm{d}\omega~\rho_D(\omega) \rho_{P(\mbf{q})}(\omega), 
\end{align}
\tb{$\Delta^2 = N_{nn}\braket{U_{nn}^2}$, $N_{nn}$ is the number of nearest neighbor strongly coupled molecules,  and $\braket{U_{nn}^2}$ is the mean squared fluctuation of the intermolecular transition-matrix element $[V_\t{MM}]_{ij}$ (where $i$ and $j$ are nearest-neighbors). Note that our definition of $\Delta^2$ differs slightly from Ref. \cite{li2022polariton}, as we perform an incoherent sum over the possible final states. This follows from our assumption that the molecular dipole vectors are randomly distributed and uncorrelated (so there is no interference between $D \leftarrow P(\mbf{q})$ pathways mediated by transition-matrix elements $[V_{\t{MM}}]_{ij}$ with fixed molecule $i$ and its nearest neighbors $j$).}

\par \tb{We will obtain an upper bound estimate for $ k_{D\leftarrow P(\mbf{q})}^{\t{dd}, M}$ assuming $\rho_{P(\mbf{q})}$ and $\rho_D(\omega)$ are centered at $E_M$ and have equal linewidth $\gamma_M$. This gives the upper bound $S_{P(\mbf{q})}^0 > S_{P(\mbf{q})}$. We will employ lineshapes for the weakly coupled modes given by $\rho_D(\omega) = N(\gamma_M,\omega_M)(\gamma_M/\pi)\times 1/[(\omega-\omega_M)^2+\gamma_M^2]$, where $N(\gamma_M,\omega_M) $ is a normalization constant and $\omega > 0$. To estimate $\Delta^2$ in the absence of knowledge of the magnitude of the dipole-dipole coupling fluctuations, we assume $U_{nn}$ is equal to the isotropically averaged magnitude of (free space) dipole-dipole interactions for a system with $N_{nn} = 6$ with mean distance $a$ and dipole magnitude $\mu$}
\begin{align}
\braket{\Delta^2} \approx \frac{2}{3} N_{nn} \frac{\mu^4}{16\pi^2 \epsilon^2  a^6\hbar^2},
\end{align}
\tb{where $\epsilon = n^2 \epsilon_0$ and $n$ is the (optical) index of refraction of the medium. Using our estimate for $U_{nn}$, the upper bound $S_{P(\mbf{q})}^0$ and $P_P^{\t{M}}(q) \approx 1$ we find that}
\begin{align}
  k_{D\leftarrow P(\mbf{q})}^{\t{dd}, M} <   \frac{\mu^4}{2\pi \epsilon^2  a^6\hbar^2} S_{P(\mbf{q})}^0
\end{align}
\tb{For an intermolecular distance of $ a = 1~\t{nm}$,  $\mu = 1~\t{D}$, $n = 1.33$ (corresponding to methanol, the solvent employed in Refs. \cite{thomas2016,thomas2019}), and $1/\gamma_M = 1~\t{ps}$ (approximately 4.1 meV/$\hbar$), we find $1/k_{D\leftarrow P(\mbf{q})}^{\t{dd}, M} > 16 ~\t{ps}$. Under similar conditions where $P_M^P(\mbf{q}) \rightarrow 1$, we find $1/k_{D\leftarrow P(\mbf{q})} ^{\t{SB}, M} \approx 1~ \t{ps}$.}

\tb{The given estimates support the view that inelastic scattering induced by fluctuations of the dipole-dipole interaction (between molecular normal modes involved in polariton formation) provides, in general, a negligible change to the rates obtained from Eq. \ref{eq:gamma_inel}. The case where the intermolecular dephasing pathway would be expected to be most relevant is that with the shortest mean intersolute distance $a = 0.7 ~\t{nm}$ (corresponding to the atypically large 5 mol/L concentration). Nevertheless, even in this case, when we take into account that $\gamma_M = 3.97 ~\t{meV}/\hbar$, $\gamma_C = 3.47~\t{meV}/\hbar$ and $E_M - E_{P(\mbf{q})} \approx 5 ~\t{meV}$ at LP mode with $q = q_{\t{max}}^{\t{L}}$ (which has 89.7$\%$ molecular weight and 10.3$\%$ photonic weight), we find the contribution of  $k_{D\leftarrow P(\mbf{q})}^{\t{dd}, M}$ to the inelastic scattering rate is weaker by a factor of O(10) relative to the rate given by Eq. \ref{eq:gamma_inel} (assuming $\mu = 1~\t{D}$ and with spectral overlap obtained from analogous distributions for the weakly coupled and polariton modes with linewidths $\gamma_M$ and $P_M^{P}(\mbf{q}) \gamma_M + P_C^{P}(\mbf{q}) \gamma_C$, respectively).}

\par \tb{Importantly, in most molecular systems where a solute interacts strongly with the infrared EM field, $a$ tends to be greater than 1 nm (i.e., concentration less than $\approx 1.65~\t{mol/L}$), and the transition dipole moments are also smaller than employed in our estimates, i.e., $\mu < 1~\t{D}$. The situation could very well be different in close-packed systems with short mean intermolecular distances $a \ll 1~ \t{nm}$ and large transition dipoles $ \mu \gg 1~\t{D}$. The former scenario is more common in pure solvent strong vibrational strong coupling or organic electronic exciton strong coupling in molecular crystals \cite{kena-cohen_strong_2008-1}. However, even in these systems, the Rabi splitting tends to be significantly larger than considered in our analysis due to the high molecular density, and thus, the spectral overlap between polaritons with small photonic content and weakly coupled molecular modes is also likely smaller than in the upper bound computed above.}

\bibliography{lib.bib}

\end{document}